\documentclass[11pt]{article}

\usepackage[numbers]{natbib}
\usepackage{authblk}

\usepackage{comment}
\usepackage[english]{babel}
\usepackage[letterpaper,top=2cm,bottom=2cm,left=3cm,right=3cm,marginparwidth=1.75cm]{geometry}

\usepackage{amsmath,amssymb}
\usepackage{graphicx}
\usepackage{float}
\usepackage{color}
\usepackage{enumitem}
\usepackage{siunitx}
\usepackage{subcaption}
\usepackage{tikz}
\usepackage{pgfplots}
\usepackage{bm}
\usepackage[colorinlistoftodos]{todonotes}

\usepackage{sectsty,booktabs,fixltx2e,setspace}
\usepackage[toc]{appendix}
\usepackage[colorlinks=true,allcolors=blue]{hyperref}
\usepackage{multicol}
\usepackage{blindtext}
\usepackage[export]{adjustbox}

\tikzset{every picture/.style={font issue=\footnotesize},
         font issue/.style={execute at begin picture={#1\selectfont}}
         }

\DeclareMathOperator{\bfn}{\mathbf{n}}
\DeclareMathOperator{\sgn}{\text{sgn}}

\title{A predictor-corrector scheme for approximating signed distances using finite element methods}

\author[1,2]{Amina El Bachari}
\author[3]{Johann Rannou}
\author[2]{Vladislav A. Yastrebov}
\author[2]{Pierre Kerfriden}
\author[1]{Susanne Claus}

\affil[1]{DTIS, ONERA, Université Paris-Saclay, 91120 Palaiseau, France}
\affil[2]{Centre de Matériaux, MINES Paris-PSL, CNRS UMR 7633, 78000 Versailles, France}
\affil[3]{DMAS, ONERA, Université Paris-Saclay, 92320 Châtillon, France}

\begin{document}

\maketitle
\begin{abstract}
In this article, we introduce a finite element method designed for the robust computation of approximate signed distance functions to arbitrary boundaries in two and three dimensions. Our method employs a novel prediction-correction approach, involving first the solution of a linear diffusion-based prediction problem, followed by a nonlinear minimization-based correction problem associated with the Eikonal equation. The prediction step efficiently generates a suitable initial guess, significantly facilitating convergence of the nonlinear correction step. \\
A key strength of our approach is its ability to handle complex interfaces and initial level set functions with arbitrary steep or flat regions, a notable challenge for existing techniques. Through several representative examples, including classical geometries and more complex shapes such as star domains and three-dimensional tori, we demonstrate the accuracy, efficiency, and robustness of the method, validating its broad applicability for reinitializing diverse level set functions.
\end{abstract}

\section{Introduction}
The computation of signed distance functions (SDFs) plays a critical role in numerous computational methods, particularly within computational mechanics, computer graphics, and numerical simulations involving interface capturing and tracking \cite{osher2003level,sethian1999level}. Accurate and efficient methods to compute SDFs are crucial for applications such as shape optimization \cite{CRMATH_2002__334_12_1125_0,CRMATH_2023__361_G8_1267_0}, multi-phase flow \cite{Sussman1994, henri2021, claus2019cutfem} or image processing \cite{Tsai2003,reinit_DRLSE}. The signed distance function is typically defined as the solution of the Eikonal equation, a nonlinear partial differential equation whose numerical solution poses significant computational challenges \cite{sethian1996fast}.

Traditional methods to compute SDFs generally fall into two main categories: geometric algorithms such as fast marching methods \cite{chopp1993,sethian1999level,FMsiam1999,FMkimmel1998} and fast sweeping methods \cite{FastSweepingTsai,FastSweepingZhao,FSqian2007} which can be highly efficient but are challenging to parallelize \cite{Bernacki2015}, extend to unstructered grids \cite{GFMM},\cite{FMM_TetrahedralDomain} or to higher-order geometric representations. PDE-based methods, which solve variants of the Eikonal equation directly using finite element or finite difference discretizations, offer greater flexibility for complex domains but require careful treatment to ensure stability and accuracy \cite{osher2003level}.

In this article, we propose a PDE-based finite element approach characterized by a novel prediction-correction strategy. Our method begins with a linear diffusion-based prediction step, efficiently providing a robust initial approximation of the SDF. This is subsequently refined through a nonlinear correction step formulated as a minimization problem derived from the Eikonal equation. This minimization based formulation of the Eikonal equation has first been used in \cite{reinit_DRLSE} for the reinitialization of a level set function. As the original method struggles in regions with a gradient norm of the level set function close to zero several modifications of the original problem have been proposed in~\cite{reinit_Basting,xue_new_2021} which modify the equations for small norms of the gradient of the signed distance functions. However, these changes are not consistent with the original problem of solving the Eikonal equation. In this article, we propose to leave the minimization based problem unchanged. Instead, we focus on formulating a prediction problem to provide an ideal initial guess to the minimization based formulation of the Eikonal problem. We will demonstrate that this yields a robust predictor-corrector method that can handle initial functions with arbitrary steep and flat regions for complex two- and three-dimensional geometries. 

We demonstrate the use of our method in the context of meshes that fit to the boundary of the complex geometry to which we want to compute the signed distance function as well as the case in which the boundary is immersed in a background finite element grid. In the case of immersed geometries, we reconstruct the interface and enforce Dirichlet conditions weakly on the unfitted interface using cut integration~\cite{CutFEM2014}. The implementation of our method is done in FEniCSx~\cite{Baratta2023_Dolfinx} and CutFEMx~\cite{CutFEMx}.

We validate our method through extensive numerical experimentation, including classical benchmarks such as circles and tori, as well as challenging test cases featuring noisy data and non-smooth boundaries. Our results highlight the accuracy, computational efficiency, and robustness of the approach, demonstrating its potential as a versatile and reliable tool for reinitializing level set functions in diverse applications within computational mechanics and engineering.

The remainder of this article is organized as follows. In Section~\ref{ls_method_intro}, we introduce the mathematical formulation of the signed distance function and corresponding signed distance problem. In Section~\ref{Elliptic_method_intro} we discuss the minimization-based Eikonal problem and introduce its finite element formulation. Section~\ref{PCscheme} presents the prediction-correction algorithm and its finite element discretization. In Section~\ref{Numerical_study}, we validate our predictor-corrector scheme on a range of numerical experiments in both two- and three-dimensions. 

\section{Level set method}\label{ls_method_intro}
\subsection{Level sets and signed distance functions.}
Let's consider fixed domain $D\subset\mathbb{R}^{n}$ containing implicitly defined domain $\Omega$. A level set function, $\phi:D\rightarrow\mathbb{R}$, is used to represent $\Omega\subset D$ such that:\par
\begin{equation}\label{eq:7}
\begin{cases}
\phi(x)<0 & \text{ if }x\in\Omega, \\
\phi(x)=0 & \text{ if }x\in \Gamma, \\
\phi(x)>0 & \text{ if }x\in D\setminus\left\{{\Omega\cup\Gamma}\right\}
\end{cases}
\end{equation}
with $\Gamma = \left\{x\in D \text{ such that } x\in\partial\Omega \setminus\partial D\right\}$.\\
A level set function with signed distance property with respect to $\phi(x)=0$ is defined as:
\begin{equation}
\begin{aligned}
\phi(x) =&
\begin{cases}
-d\left(x,\Gamma\right) & \text{ if }x\in\Omega,\\
d\left(x,\Gamma\right) & \text{ if }x\in D\setminus\left\{{\Omega\cup\Gamma}\right\},
\end{cases}
\end{aligned}
\end{equation}
where $d$ is the euclidean distance function defined as: 
\begin{equation}
\begin{aligned}
d\left(x,\Gamma\right)=\underset{y\in\Gamma}{\inf}d\left(x,y\right).
\end{aligned}
\end{equation}
If the domain $\Omega \subset \mathbb{R}^{n}$ has a piecewise smooth boundary, then the signed distance function to  domain boundary $\Gamma$ is differentiable almost everywhere, and the equivalent signed distance property (SDP) holds 
\begin{equation}\label{eq:signed_distance}
\left|\nabla\phi\right|= 1\text{, almost every where in } D,
\end{equation}
where \( |\cdot| \) denotes the Euclidean norm, defined by
\[
|x| = \left( \sum_{i=1}^{n} x_i^2 \right)^{1/2},
\]
with \( x = (x_1, x_2, \ldots, x_n) \in \mathbb{R}^n \). \\
The normal to the contour lines of locally smooth level set function is given by
\begin{equation}\label{eq:signed_distance_normal}
\bfn_{\phi}=\frac{\nabla\phi}{\left|\nabla\phi\right|},
\end{equation}
which can be simplify as $\bfn_{\phi}=\nabla\phi$, almost every where, for a level set function with signed distance property.

\subsection{Need for approximate redistancing}
In general, the level set method is employed in the context of moving boundary problems. In this framework, an advection equation is used to advect the domain boundary $\Gamma$ in the normal direction $ \theta = v \bfn_{\phi}$. Regardless of the method used for the advection of the level set,
the signed distance property $|\nabla \phi|= 1$ is not preserved. 
Nevertheless, maintaining the signed distance property is essential, as it greatly facilitates the following aspects: 
\begin{enumerate}[label=(\roman*)]
        \item to compute normals on $\Gamma$, $\bfn_{\phi} = \nabla \phi/|\nabla \phi|$, which becomes inaccurate when $|\nabla \phi| \rightarrow 0$;
        \item for the level set advection, because the term $\theta \cdot \nabla \phi=v\left|\nabla\phi\right|$  becomes difficult to stabilize for steep $\phi$;
        \item to reconstruct the interface by linear interpolation, which becomes inaccurate for flat $\phi$ as too small change in $\phi$ values yields a large difference in determining the location of $\Gamma$.
    \end{enumerate}
The use of the level set method for numerous moving boundary problem (shape optimization, fluid simulation, propagation phenomena) has led to active research on reinitialization methods to address the diverse needs associated with each application.

The need for redistancing extends beyond moving boundary value problems. For instance, solving partial differential equations on domains derived from the semantic segmentation of an image can be made more efficient by redefining the domain's contour using the zero level-set, rather than relying on pixel or voxel boundaries (see for example \cite{clausCutFEMContact}). In this context, the level-set is not introduced as a distance function but rather as a function that is typically constant inside and outside of the domain, and piece-wise polynomial in a band of interface elements. Consequently, the resulting level-set does not possess signed distance properties. Another example arises in mesh generation, where signed distance functions are used to represent the geometry and to guide the construction of the adaptive meshes (see \cite{Molino2003}). More broadly, signed distance functions are widely employed in computer graphics (see \cite{Erleben2008}).

\subsection{Signed Distance Problem}\label{Reinit_method_intro}
Starting from a level set function $\phi_{0}$ that does not have the signed distance property, we seek to determine the corresponding signed distance function denoted $\phi$ that is satisfying the following properties:
\begin{align}\label{eq:reinitMethod}
\begin{cases}
\left|\nabla\phi\right| & \!\!\!\! = 1 \qquad \quad \text{ in } D, \\
\phi(x) & \!\!\!\!= 0 \qquad \quad \text{ on } \Gamma, \\
\text{sgn}\left(\phi\right) & \!\!\!\!  =\text{sgn}\left(\phi_{0}\right) \text{ in } D\setminus\Gamma.
\end{cases}
\end{align}
To solve this problem, reinitialization methods are used.
A wide range of methods are proposed in the literature; in the 
following section, we will present the used  minimization based method.

\section{Minimization based reinitialization}\label{Elliptic_method_intro}
\subsection{Least squares minimization}
Our redistancing method is inspired by Basting and Kuzmin \cite{reinit_Basting} who formulated the Eikonal equation, $\left|\nabla\phi\right| = 1$, as a least squares minimization problem 
\begin{equation}\label{minproblem}
   \min \left( \int_{D} \frac{1}{2} \left(\left|\nabla\phi\right|-1\right)^{2}\, dx \right) .
\end{equation}
Setting
\begin{equation}
    \mathcal{F}(\phi) = \left( \int_{D} \frac{1}{2} \left(\left|\nabla\phi\right|-1\right)^{2}\text{ }dx \right) 
\end{equation}
and taking the Gâteaux derivative of $\mathcal{F}(\phi)$ in direction $\psi$, we obtain 
\begin{align} 
d\mathcal{F}(\phi;\psi) &= \int_{D} \frac{1}{2} 2 (|\nabla \phi| - 1) \lim_{\tau \rightarrow 0} \left( \frac{|\nabla (\phi + \tau \psi)| - |\nabla \phi|}{\tau} \right) \, dx  \\
&= \int_{D} (|\nabla \phi| - 1)  \frac{\nabla \phi}{|\nabla \phi|} \nabla \psi \, dx \\
&= \int_{D} \left(\nabla\phi-\frac{\nabla\phi}{\left|\nabla\phi\right|}\right) \nabla \psi \, dx 
\end{align}
And the minimization problem becomes 
\begin{equation}
\int_{D} \left(\nabla\phi-\frac{\nabla\phi}{\left|\nabla\phi\right|}\right) \nabla \psi \, dx = 0.
\end{equation}
We can derive the strong form by integration by parts
\begin{equation}
- \int_{D} \nabla \cdot \left(\nabla\phi-\frac{\nabla\phi}{\left|\nabla\phi\right|}\right) \psi \, dx  +  \int_{\partial D} \left( \left(\nabla\phi-\frac{\nabla\phi}{\left|\nabla\phi\right|}\right) \cdot \bfn \right) \psi \, ds = 0,
\end{equation}
and setting 
\begin{equation}
\left(\nabla\phi-\frac{\nabla\phi}{\left|\nabla\phi\right|}\right)\cdot \bfn =0\text{ on }\partial D,
\end{equation}
which yields 
\begin{align}\label{equation_Corrector}
\begin{cases}
-\nabla\cdot\left(\nabla\phi-\frac{\nabla\phi}{\left|\nabla\phi\right|}\right) &= 0 \quad \text{in } D,\\
\left(\nabla\phi-\frac{\nabla\phi}{\left|\nabla\phi\right|}\right)\cdot \bfn &= 0 \quad \text{on } \partial D.
\end{cases}
\end{align}
Here, $\bfn$ denotes the outward normal to the boundary $\partial D$. 
To obtain the distance to the boundary defined by $\Gamma$, we impose the following Dirichlet condition 
\begin{equation}
\phi = 0 \mbox{ on } \Gamma. 
\end{equation}
The solution to this non-linear problem is the unsigned distance to boundary $\Gamma$. This problem is very stiff and therefore very difficult to solve and does not have any notion of a sign of $\phi$. \\ These two challenges can be overcome by reformulating \eqref{equation_Corrector} as an iterative scheme with an initial signed solution $\phi^0$ and $\phi^0 = 0$ on $\Gamma$. There are several options to formulate an iterative scheme. One possibility is to add an artificial time term $\frac{\partial \phi}{\partial \tau}$ to \eqref{equation_Corrector} and/or to use a fixed point iteration scheme. In this article, we use the most straightforward iterative scheme given by 
\begin{align}\label{equation_Corrector_iterative}
\begin{cases}
-\nabla\cdot \nabla\phi^{n+1} &= -\nabla\cdot \dfrac{\nabla\phi^{n}}{\left|\nabla\phi^n\right|} \quad \text{ in }D\\
\nabla\phi^{n+1} \cdot \bfn&= \dfrac{\nabla\phi^n}{\left|\nabla\phi^{n}\right|} \cdot \bfn \quad \quad \text{ on }\partial D,
\end{cases}
\end{align}
which is a fixed point Picard iteration scheme that linearizes the problem. Here, $n$ denotes the $n$-th iterate. \\
All these iterative schemes have one major drawback:
\begin{equation}
\text{If } |\nabla \phi^n| \rightarrow 0 \text{, then } \frac{1}{|\nabla \phi^n|} \rightarrow \infty. 
\end{equation}
If we do not pose any conditions on our initial function $\phi^0$ then none of the iterative schemes for problem \eqref{equation_Corrector} will converge for flat areas of $\phi^0$. One way to prevent this  is to reformulate the original problem as suggested by \cite{reinit_Basting,xue_new_2021, AdamsEllipticPDEReinitDG} and detailed in the next Section. In \cite{XUE2021}, it is proposed to solve the Laplace equation in a neighborhood of the interface. The conditions on this neighborhood are chosen so as to preserve the sign of the initial level set. Nevertheless, this method cannot be extended to the entire domain $D$. In this article, however, we propose to address this shortcoming of the method by carefully designing a predictor problem which we solve to obtain an optimal initial guess $\phi^0$. Before discussing our method, we will briefly review the reformulations of \cite{reinit_Basting, AdamsEllipticPDEReinitDG} and show why we felt the need to design a predictor problem instead. 

\subsection{Modified minimization and non-linear diffusion formulation}
In order to prevent $\frac{1}{|\nabla \phi|} \rightarrow \infty $, \cite{reinit_Basting} and \cite{AdamsEllipticPDEReinitDG} propose to use another function, $p(|\nabla \phi|)$, in  
\begin{equation}\label{cost_term}
    \mathcal{F}\left(p,\phi\right)=\int_{\Omega}p\left(\left|\nabla\phi\right|\right)\, dx.
\end{equation}
For which the Gâteaux derivative is given by
\begin{align} 
d\mathcal{F}(\phi;\psi) &= \int_{D} \frac{p'(|\nabla \phi|)}{|\nabla \phi|} \nabla \phi \nabla \psi \, dx.
\end{align}
Looking for the zero of this Gâteaux derivative can be interpreted as a non-linear diffusion problem 
\begin{align} 
\int_{D} \widetilde{d}\left(\left|\nabla\phi\right|\right) \nabla \phi \nabla \psi \, dx =0 ,
\end{align}
where the non-linear diffusion term is given by
\begin{equation}
\widetilde{d}\left(\left|\nabla\phi\right|\right) := \frac{p'(|\nabla \phi|)}{|\nabla \phi|} .
\end{equation}
The strong form for this non-linear diffusion problem becomes 
\begin{align} 
\begin{cases}
- \nabla \cdot \left(\widetilde{d}\left(\left|\nabla\phi\right|\right) \nabla \phi \right) &= 0 \, \text{ in } D, \\
\left(\widetilde{d}\left(\left|\nabla\phi\right|\right) \nabla \phi \right) \cdot n &= 0 \, \text{ on } \partial D .
\end{cases}
\end{align}
Basting et al. \cite{reinit_Basting} introduced the following double well potential inspired function for $\left|\nabla\phi\right|\leq1$
\begin{equation}
\label{equ: potential 1}
    p\left(\left|\nabla\phi\right|\right):=\begin{cases}
\frac{1}{2}\left(\left|\nabla\phi\right|-1\right)^{2} & \text{if }\left|\nabla\phi\right|>1, \\
\frac{1}{2}\left|\nabla\phi\right|^{2}\left(\left|\nabla\phi\right|-1\right)^{2} & \text{if }\left|\nabla\phi\right|\leq1 .
\end{cases}
\end{equation}
and keeping the original function for $\left|\nabla\phi\right|>1$. 
The diffusion term for this function becomes
\begin{equation}\label{diff_2}
    \widetilde{d}\left(\left|\nabla\phi\right|\right):=\begin{cases}
1-\frac{1}{\left|\nabla\phi\right|} & \text{if }\left|\nabla\phi\right|>1,\\
1-(3\left|\nabla\phi\right|-2\left|\nabla\phi\right|^{2}) & \text{if }\left|\nabla\phi\right|\leq1.
\end{cases}
\end{equation}
Adams et al. \cite{AdamsEllipticPDEReinitDG} pointed out that for $|\nabla \phi| < 0.5$ this diffusion term becomes positive meaning that an already flat functional profile is pushed towards $\left|\nabla \phi\right|=0$ in any iterative scheme. In the original formulation the diffusion term is positive for $|\nabla \phi| > 1$, which causes the function profile to flatten and it is negative for $|\nabla \phi| < 1$ which causes the function profile to steepen through the iterations.  Adams et al. \cite{AdamsEllipticPDEReinitDG} suggested the following modification of the function 
\begin{equation}
\label{equ: potential 2}
    p\left(\left|\nabla\phi\right|\right):=\begin{cases}
\frac{1}{2}\left(1-\frac{1}{\left|\nabla\phi\right|}\right) & \text{if }\left|\nabla\phi\right|>1\\
\frac{\left(\left|\nabla\phi\right|\right)^{3}}{3}-\frac{\left(\left|\nabla\phi\right|\right)^{2}}{2}+\frac{1}{6} & \text{if }\left|\nabla\phi\right|\leq1.
\end{cases}
\end{equation}
and the corresponding diffusion term becomes 
\begin{equation}\label{diff_3}
    \widetilde{d}\left(\left|\nabla\phi\right|\right):=\begin{cases}
1-\frac{1}{\left|\nabla\phi\right|} & \text{if }\left|\nabla\phi\right|>1\\
1-(2-\left|\nabla\phi\right|) & \text{if }\left|\nabla\phi\right|\leq1.
\end{cases}
\end{equation}
To solve these non-linear diffusion problems, \cite{reinit_Basting}, \cite{AdamsEllipticPDEReinitDG} solve this problem by Picard iteration, which for the strong form can be written as
\begin{equation}
\label{equ: picard diffusion}
\begin{aligned}
   - \nabla \cdot  \nabla \phi^{n+1}  &= - \nabla \cdot \left(\widetilde{d}^*\left(\left|\nabla\phi^n\right|\right) \nabla \phi^n \right)& \, \text{ in } D ,\\
 \nabla \phi^{n+1}  \cdot \bfn &= \left(\widetilde{d}^*\left(\left|\nabla\phi^{n}\right|\right) \nabla \phi^n \right) \cdot \bfn & \, \text{ on } \partial D  ,
\end{aligned}
\end{equation}
Here,
\begin{align}
\label{d*1}
\widetilde{d}^*\left(\left|\nabla\phi\right|\right) =  \frac{1}{\left|\nabla\phi\right|}
\end{align}
in the original problem \eqref{equation_Corrector} and for $|\nabla \phi| > 1$ for \eqref{diff_2} and \eqref{diff_3}. And for $|\nabla \phi| \leq 1$, \eqref{diff_2} yields
\begin{equation}
\widetilde{d}^*\left(\left|\nabla\phi\right|\right)= 
(3\left|\nabla\phi\right|-2\left|\nabla\phi\right|^{2})
\end{equation}
and \eqref{diff_3} yields 
\begin{equation}
\widetilde{d}^*\left(\left|\nabla\phi\right|\right)= 
(2-\left|\nabla\phi\right|).
\end{equation}

\subsection{Finite element formulation for the non-linear diffusion problem}
Let $\phi^0: D \rightarrow \mathbb{R}$ be a function who decomposes our domain $D$ into two parts:  a negative part $\Omega^-$ and a positive part $\Omega^+$ given by the sign of the function $\phi^0$, and an interface between  $\Omega^-$  and $\Omega^+$, denoted by $\Gamma$,  given by the zero contour line of $\phi^0$. This means 
\begin{align}
    \phi^0(x) &> 0 \quad \text{ for } x \in \Omega^+, \\
    \phi^0(x) &= 0 \quad \text{ for } x \in \Gamma, \\
    \phi^0(x) &< 0 \quad \text{ for } x \in \Omega^-.
\end{align}
We do not make any further assumptions on $\phi^0$.  \\
Let us now presume that $D$ is decomposed into mesh elements whose faces coincide with $\Gamma$. We denote this mesh by $\mathcal{K}_h$ and the faces that coincide with $\Gamma$ by $\Gamma_h$ as well as the discretized domain by $D_h = \Omega^+_h \cup \Omega^-_h$. In later sections in this article, we will discuss how this scheme can be extended to meshes that do not coincide with $\Gamma_h$. \\
Furthermore let $V_h^0$ denote the continuous piecewise linear finite element space which includes the zero Dirichlet boundary condition on $\phi$ along $\Gamma_h$, i.e. 
\begin{equation}\label{spaceVh_0}
  V_{h}^0:=\left\{ v\in H^1\left(D\right)\cap C^{0}\left(D\right) \mid v_{\mid K}\in \mathbb{P}_{1}\left(K\right) \text{ for all }K\in\mathcal{K}_{\;h}, \, v_{\mid F} = 0 \text{ for all }F\in \Gamma_{h} \right\}.  
\end{equation}
\\
Our finite element formulation to solve the non-linear diffusion problem \eqref{equ: picard diffusion} then reads: For $n = 0 , ... ,N-1$, find $\phi_h^{n+1} \in V_h^0$ such that for all $v_h \in V_h^0$ 
\begin{equation}
\label{equ: corrector}
    \int_{D_h} \nabla \phi_h^{n+1} \nabla v_h \, dx = \int_{D_h} \left(\widetilde{d}^*\left(\left|\nabla\phi_h^n\right|\right) \nabla \phi_h^n \right) \nabla v_h \, dx . 
\end{equation}
Note that the Neumann boundary condition from \eqref{equ: picard diffusion} is implicitely enforced through first integrating by parts and then setting the boundary term to zero. And the Dirichlet condition is strongly enforced through the finite element space. The number of iterations, $N$, is either fixed or we are iterating until

\begin{equation}
   \left\Vert \left( |\nabla\phi^{n+1}|-1\right) \right\Vert_{L^{2}\left(D\right)} <\varepsilon
\end{equation}

\begin{equation}
   \left| \left\Vert \left( |\nabla\phi^{n+1}|-1\right) \right\Vert_{L^{2}\left(D\right)} - \left\Vert \left( |\nabla\phi^{n}|-1\right) \right\Vert_{L^{2}\left(D\right)} \right| <\varepsilon
\end{equation}
with a small positive real value for $\varepsilon$.

\subsubsection{Test case for non-linear diffusion problem}
\begin{figure}[H]
\centering
    \begin{subfigure}[b]{0.23\linewidth}\centering
         \includegraphics[width=0.9\textwidth]{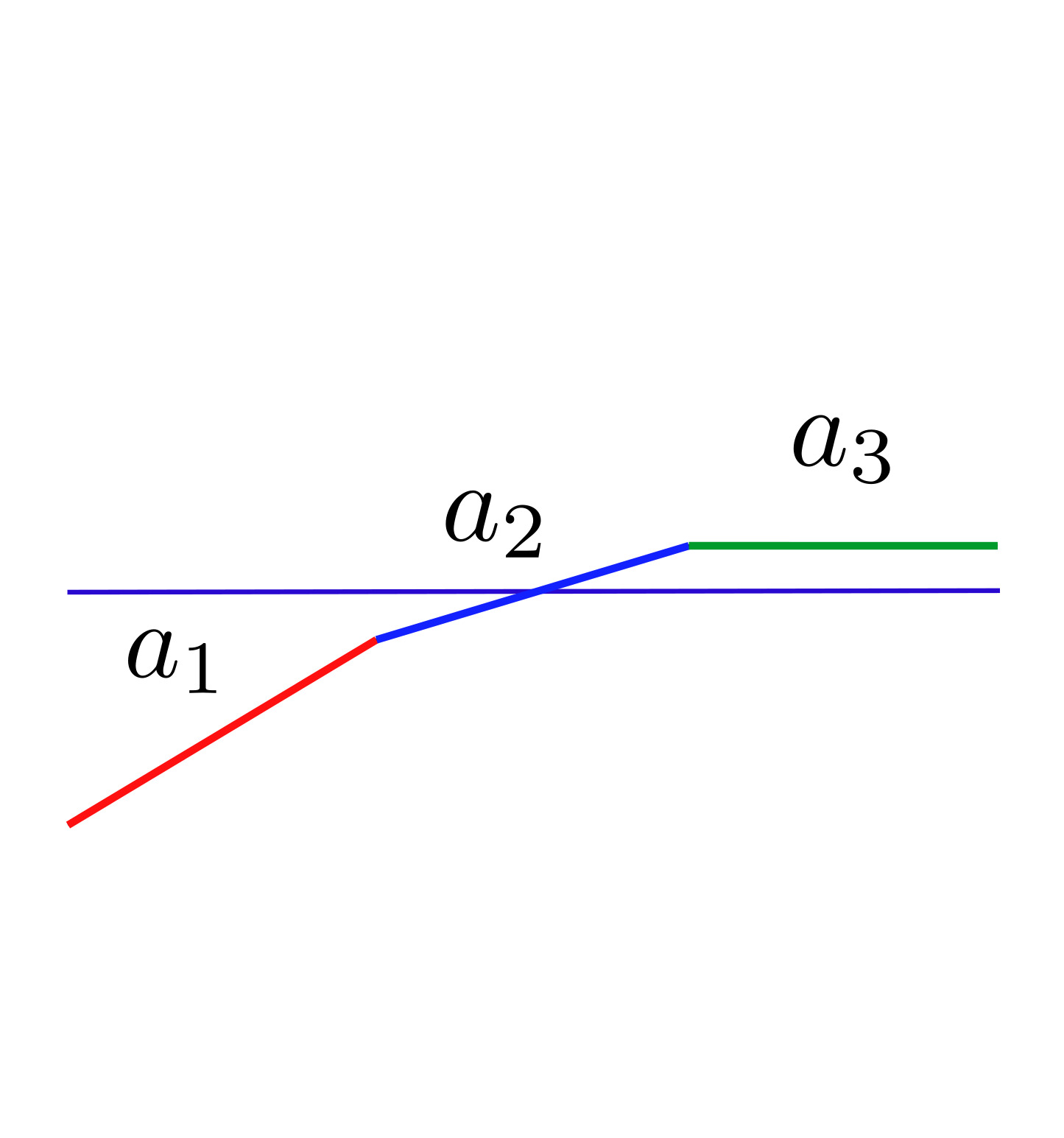}
          \caption{}\label{ComparaisonInit}
    \end{subfigure}
    \begin{subfigure}[b]{0.23\linewidth}\centering
        \includegraphics[width=0.9\textwidth]{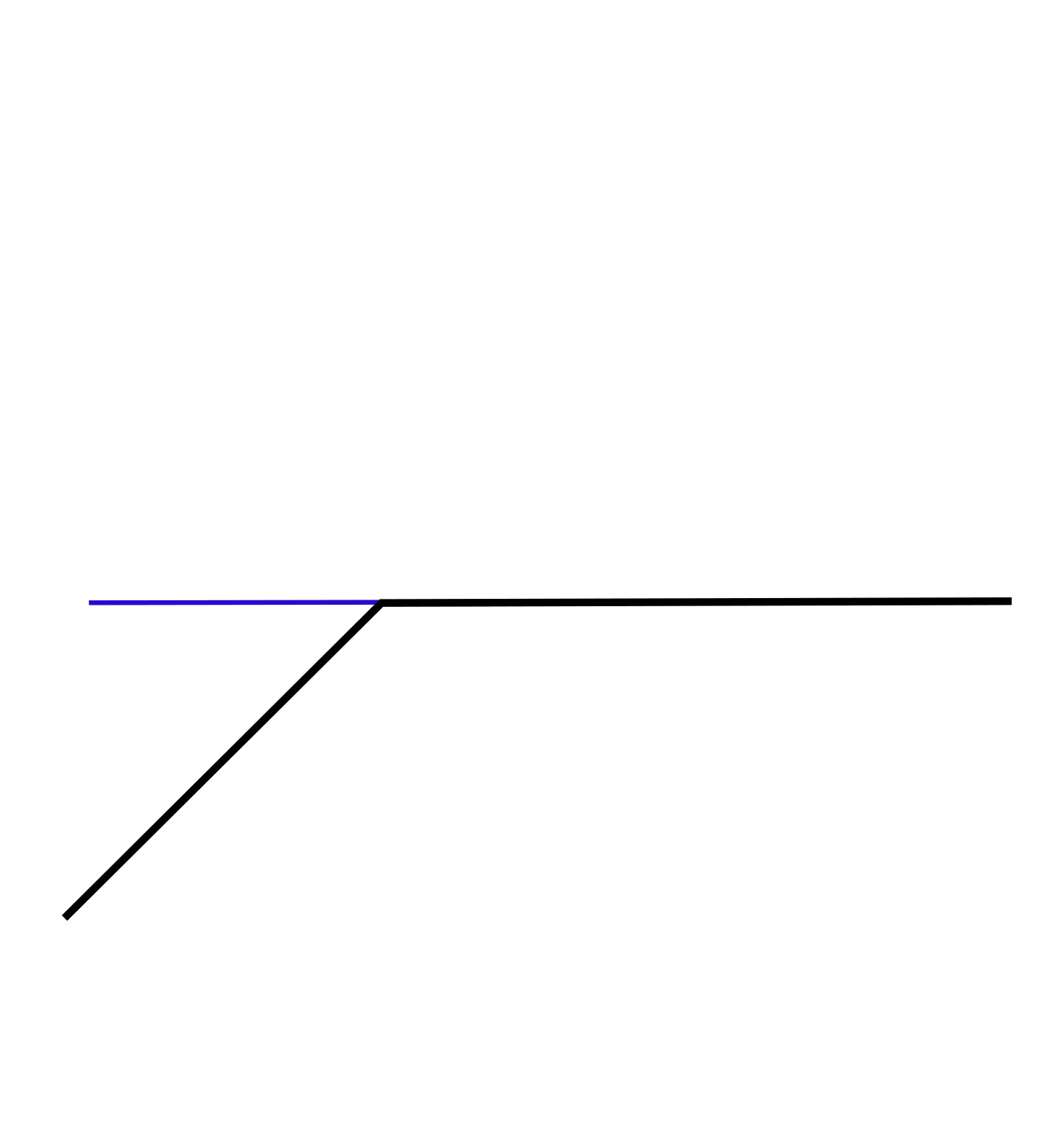}
        \caption{}\label{Kuzmin}
    \end{subfigure}
    \begin{subfigure}[b]{0.23\linewidth}\centering
         \includegraphics[width=0.9\textwidth]{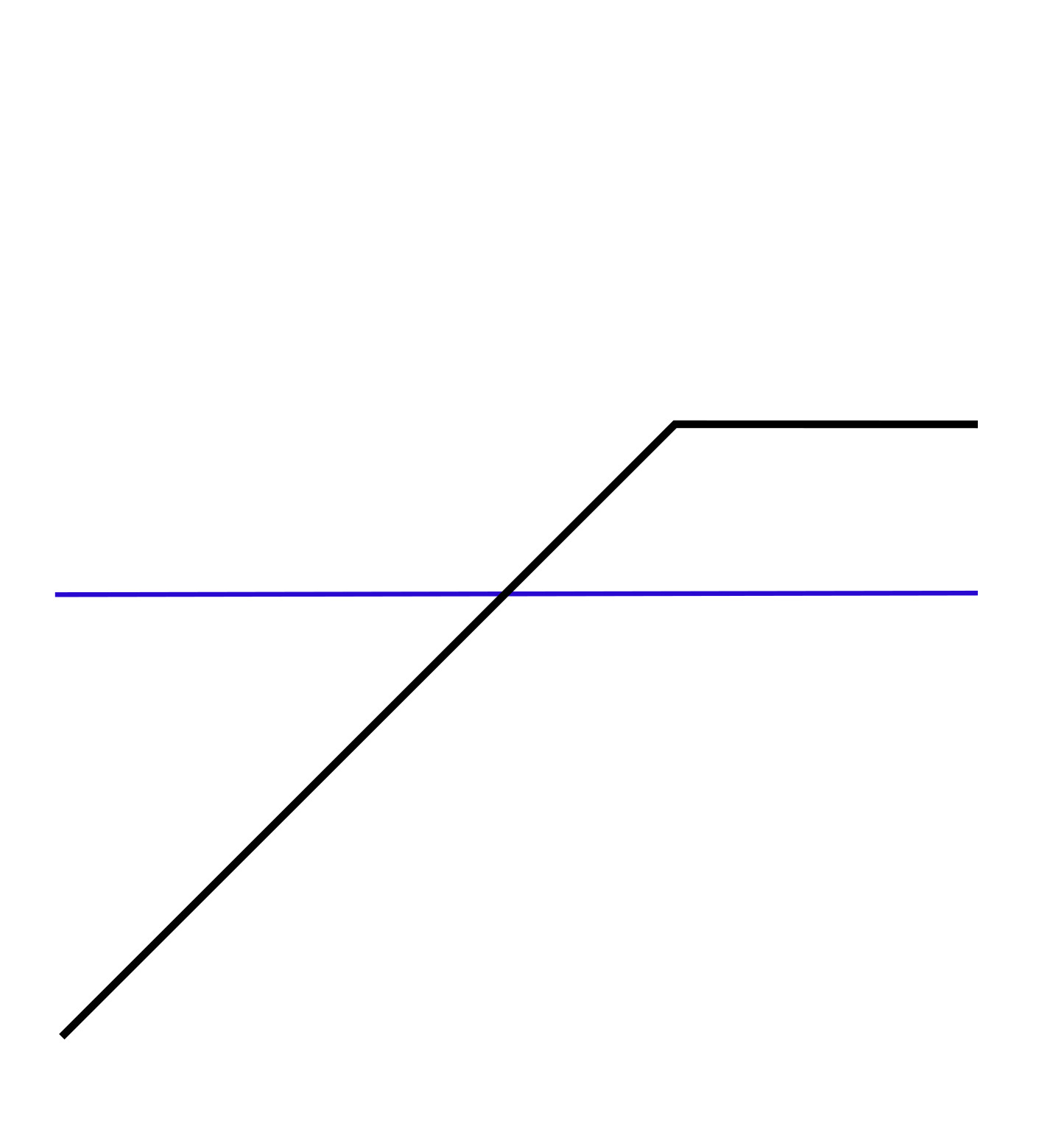}
        \caption{}\label{Adams}
    \end{subfigure}
    \begin{subfigure}[b]{0.23\linewidth}\centering
         \includegraphics[width=0.9\textwidth]{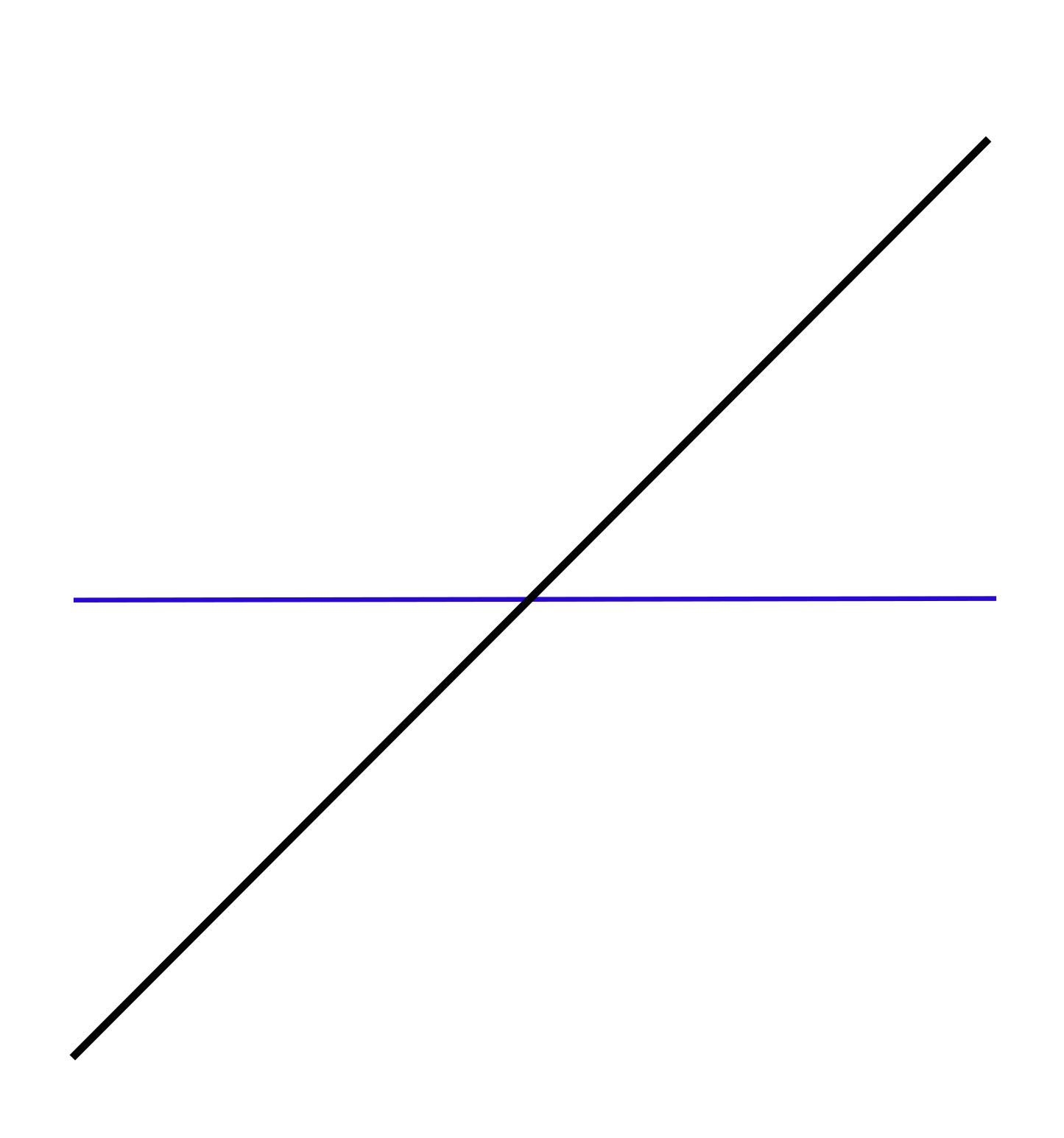}
        \caption{}\label{PC}
    \end{subfigure}
    \caption{Level set after 10 fixed point iterations, the horizontal line shows the problem domain :\subref{ComparaisonInit}Piecewise-affine initialization of the level set with specified slopes: $a_1 = 0.6$, $a_2 = 0.3$ and $a_3 = 0$; \subref{Kuzmin} Solution with \eqref{diff_2} ; \subref{Adams}Solution with \eqref{diff_3} 
    ; \subref{PC} Solution with our method.
    }
    \label{level_circle_ls}
\end{figure}

Let us assume an initial function $\phi^0$ in a domain $D = [0,1]$ defined as 
\begin{equation}
\phi^{0}(x) =
\begin{cases}
0.6x - 0.25, & \text{if } x \in \left[0, \frac{1}{3}\right] \\
0.3x - 0.15, & \text{if } x \in \left[\frac{1}{3}, \frac{2}{3}\right] \\
0.05,        & \text{if } x \in \left[\frac{2}{3}, 1\right].
\end{cases}
\end{equation}
displayed in Figure~\ref{ComparaisonInit}. This initial function features three different slopes and will be used to compare the results obtained for the different potentials proposed by Kuzmin and Basting  \cite{reinit_Basting} , Adams et al. \cite{AdamsEllipticPDEReinitDG}, and for our prediction-correction method introduced in this article. The results shown in Figure~\ref{level_circle_ls} are after 10 fixed point iterations.

First, the simulation using the function \eqref{equ: potential 1} is conducted. In the subdomain $\left[\frac{1}{3}, 1\right]$, where the level set is initialized with a function whose gradient is less than 0.5, the solution iterates towards a function with a zero gradient norm, as shown in Figure~\ref{Kuzmin}.

Secondly, the simulation using the function \eqref{equ: potential 2} is performed. The level set function converges to a signed distance function for any initial condition with a non-zero gradient norm, as shown in Figure~\ref{Adams}. However, for regions where the gradient norm is close to zero, the method does not converge to the expected signed distance function but is left unchanged. As a result, regions with a gradient norm very close to zero tend to retain their value. 

Note that in these two methods, the diffusion function is changed to avoid an infinite diffusion for small gradient norms, however, the changes are not consistent with the original problem and we are not sure of the nature of the problem that we are solving any more. The idea of minimizing the distance between the gradient norm and 1 is lost and this means that they do not always perform in the ways needed. \\ 
This is why in this article, we propose another alternative. Instead of changing the non-linear diffusion problem, we are designing a predictor problem such that we obtain a good initial guess for $\phi^0$ that removes all steep or flat areas from the function to be redistanced. We then use the original unmodified method~\eqref{equ: picard diffusion}-\eqref{d*1} to minimize the distance between $|\nabla \phi|$ and 1. The predictor-corrector scheme, proposed in this article, is able to cope with flat areas and the signed distance can be recovered for the given test case as shown in Figure~~\ref{PC}.
We will now proceed to introducing our predictor-corrector scheme.

\section{Prediction-Correction method}\label{PCscheme}
In the first Subsection, the prediction-correction method is explained and detailed. The weak imposition of Dirichlet boundary conditions using the Nitsche's method is discussed, and a simple 1D example of the prediction problem is provided. In the second Subsection, the finite element discretization is explained, and the prediction-correction scheme is discretized, with specific details provided for the interface integration.

\subsection{Predictor}\label{predictorProblem}
Let us assume we have an initial level set function, $\phi^0$ of arbitrary shape that decomposes our domain $D$ into a positive and negative part 
\begin{align}
    \Omega_{+} &:= \{ x \in D : \phi^0(x) > 0\}, \\
    \Omega_{-} &:= \{ x \in D : \phi^0(x) < 0\}, \\
    \Gamma &:=  \{ x \in D : \phi^0(x) = 0\}.
\end{align}
Furthermore, let $\sgn$ denote the sign function
\begin{equation}
\sgn(x) =
    \begin{cases}
    1 & \mbox{ for } x \in \Omega_{-}, \\
    -1 & \mbox{ for } x \in \Omega_{+}, \\
    0  & \mbox{ for } x \in \Gamma.
    \end{cases}
\end{equation}
We then formulate the following predictor problem: Find $\phi_p: D \rightarrow \mathbb{R}$, such that 
\begin{equation}\label{eq:predictive_problem}
    \begin{cases}
    -\Delta\phi_{p}  & \!\!\!\!
    = \sgn(x)\quad \text{ in } D,\\
    \phi_{p}  & \!\!\!\!
    = 0 \qquad \qquad \text{ on } \Gamma ,\\ 
    \nabla \phi_{p} \cdot \bfn & \!\!\!\!
    = \sgn(x) \quad \text{ on } \partial D .
    \end{cases}
\end{equation}
Here, we have carefully chosen $\sgn$ for the right hand side of a linear diffusion problem and as a Neumann boundary condition on $\partial D$. \\ 
The choice of $\sgn$ for the linear diffusion problem ensures the maintenance of the sign of the original function and has the same values of $|\nabla \phi_p|$ close to zero as the signed distance function, i.e. at equidistant points. This also means that there is no strong kink in the solution at $\Gamma$ as there would be if we were choosing the RHS as 1. Why this is a good first approximation for a signed distance will become even clearer in the 1D example, we give below.  \\
For the choice of the Neumann boundary condition, let us revisit the Neumann boundary condition of the problem we wish to solve~\eqref{equation_Corrector}, i.e. 
\begin{equation}
    \nabla \phi \cdot \bfn = \frac{\nabla \phi}{|\nabla \phi|} \cdot \bfn
\end{equation}
With $\bfn_{\phi}$ the normal to the contour lines of function $\phi$, the following holds 
\begin{equation}
   \frac{\nabla \phi}{|\nabla \phi|} \cdot \bfn = \bfn_{\phi} \cdot \bfn  
\end{equation}
Here, $\bfn_{\phi}$ points from $\Omega_{-}$ towards $\Omega_{+}$ and
\begin{align}
    0 \leq \bfn_{\phi} \cdot \bfn  &\leq 1 \mbox{ for } x \in \Omega_{+}, \\ 
   -1 \leq \bfn_{\phi} \cdot \bfn &\leq 0 \mbox{ for } x \in \Omega_{-}. 
\end{align}
We assume the "steepest" case 
\begin{align}
    \frac{\nabla \phi}{|\nabla \phi|} \cdot \bfn \approx \sgn(x)
\end{align}
which yields 
\begin{equation}
    \nabla \phi \cdot \bfn = \sgn(x).
\end{equation}
In addition to giving us a first approximation for the Neumann condition of the problem we want to solve, we also ensure that $\phi_p$ is not flattened near the boundary of $D$ (see the following 1D \ref{1DExample} example for more details).

\subsubsection{Illustration of prediction problem in 1D}\label{1DExample}
\begin{figure}[H]
     \centering
     \includegraphics[width=0.8\textwidth]{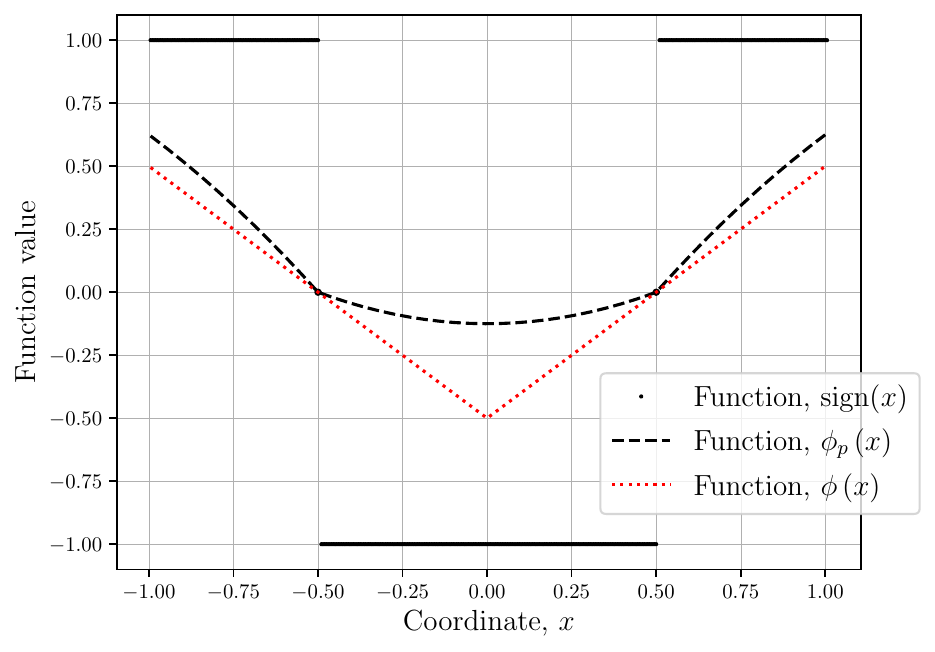}
     \caption{1D solution of the prediction-correction method, with $T = 0.5$ and $A = 1$.}
     \label{fig:predictProb1D}
\end{figure}

In one dimension the prediction problem becomes: Find a continuous function $\phi_{p} : [-A,A] \rightarrow \mathbb{R}$  such that 
\begin{align}
-\frac{\partial^2 \phi_{p}(x)}{\partial x^2} &=\sgn\left(x\right) \text{ }\forall x\in\left[-A,A\right],\\
\phi_{p}(x) &=0\text{ for }x\in\left\{ -T;T\right\}, \label{equ: interface cond 1D}\\
\frac{\partial \phi_{p}(x)}{\partial x} &=-1 \text{ at } x=-A, \label{equ: neumann -A} \\
\frac{\partial \phi_{p}(x)}{\partial x} &=1 \text{ at } x=A, \label{equ: neumann A}
\end{align}
with $T\geq0$, $T\in\left]-A,A\right[$ and $\sgn$ defined as the following step function: \\
\begin{equation}
\begin{split}
\sgn(x) :=
        \begin{cases}
        1 & \text{if }x\in\left[-A,-T\right)\cup\left(T,A\right],\\
        -1 & \text{if }x\in\left(-T,T\right),\\
        0 & \text{if }x\in\left\{ -T,T\right\}.
        \end{cases}
    \end{split}
\end{equation}
By integrating twice per interval, $[-A,-T], [-T,T],[T,A]$, and using the boundary conditions \eqref{equ: interface cond 1D}, \eqref{equ: neumann -A} and \eqref{equ: neumann A}, we obtain:
\begin{equation}\label{solution_phi_1D}
\phi_{p}(x)=\begin{cases}
-\frac{1}{2}x^{2}-\left(A+1\right)x+\frac{1}{2}T^{2}-\left(A+1\right)T & \text{ if \ensuremath{x\in\left[-A,-T\right]}} \\
\frac{1}{2}x^{2}-\frac{1}{2}T^{2} & \text{ if }\ensuremath{x\in\left[-T,T\right]}\\
-\frac{1}{2}x^{2}+\left(A+1\right)x+\frac{1}{2}T^{2}-\left(A+1\right)T & \text{ if }\ensuremath{x\in\left[T,A\right]}\\
\end{cases}
\end{equation}
Figure~(\ref{fig:predictProb1D}) shows a plot of this 1D solution (\ref{solution_phi_1D}). The solution has the same sign as the initial function and non-zero gradient norm except in the equidistant point $x=0$, which is the point that is discontinuous in the gradient in the signed distance function. \\
Note that, 
\begin{align}
\frac{\partial \phi_p}{\partial x} &= - x - (A+1) \quad \text{ for } x \in [-A,-T] \\ 
\frac{\partial \phi_p}{\partial x} &= x \quad \text{ for } x \in [-T,T] \\
\frac{\partial \phi_p}{\partial x} &= - x + (A+1) \quad \text{ for } x \in [T,A]
\end{align}
Thus 
\begin{equation}
    \frac{\nabla \phi_p}{|\nabla \phi_p|} = 
    \begin{cases}
    -1 , \quad &x \in [-A,-T)\cup(T,A] \\ 
    1 , \quad &x \in (-T,T)
    \end{cases}
\end{equation}
with $\frac{\nabla \phi_p}{|\nabla \phi_p|}$ undefined at $x=0$, exactly like the sign function. This now means in the corrector problem, for which we take \eqref{equ: corrector} 
\begin{equation}
    \int_{D_h} \nabla \phi_h \nabla v_h \, dx = \int_{D_h} \frac{\nabla \phi_p}{|\nabla \phi_p|}  \nabla v_h \, dx . 
\end{equation}
with $\nabla v_h =1$ for piecewise linear shape functions and 1D
\begin{equation}
    \int_{D_{h}} \frac{ \partial \phi_h}{\partial x} \, dx = \int_{D_{h}} \frac{\nabla \phi_p}{|\nabla \phi_p|}  \, dx . 
\end{equation}
which after integrating becomes 
\begin{equation}
\phi(x) = \frac{\nabla \phi_p}{|\nabla \phi_p|} x  - T \text{ as } \phi(-T) = 0. 
\end{equation}
This is the signed distance function 
\begin{equation}
\phi(x) = \begin{cases}
-x - T , \quad  x \in [-A,0) \\ 
x - T , \quad x \in (0,A],
\end{cases}
\end{equation}
and hence the exact solution. This shows that by solving the predictor problem followed by one fixed point iteration, we obtain the exact solution for our simple test case.  In practice, more iterations are needed, as the boundary of $\partial D$ is usually not aligned with $\bfn_{\phi}$ and we introduce numerical errors for the points with $|\nabla \phi|\approx 0$. Nevertheless it motivates our choice for this predictor problem to be a very good first initial guess for the Picard iteration scheme \eqref{equ: corrector}. 
\subsection{Brief discussion on the gradient norm close to zero}
As mentioned above the minimization based (re)-distancing method suffers from one major drawback, which is the division by the norm of the gradient. This norm of the gradient will inevitably become close to zero for a signed distance function in points that are equidistant to multiple points on the boundary. For example, for a circular boundary this is the point in the centre of the circle. For other geometries these might be points along equidistant lines or surfaces. The signed distance function is not differentiable at these points and piecewise polynomial approximations around these points can yield a gradient norm close to zero. Our predictor problem also features these "smoothened" gradient norms close to zero around equidistant points. We show below that setting a very small minimal value for the gradient norm in these cases works well in the iteration scheme. Concretely, we set $\max(|\nabla \phi|, \epsilon)$ in the denominator with $\epsilon>0$ very small. Alternatively, a narrow band approach could be used. \\

\begin{figure}[H]
\centering
    \begin{subfigure}[b]{0.8\linewidth}\centering

    \includegraphics[width=0.8\textwidth]{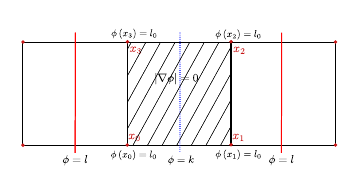}\label{2Dimension}
    \caption{}
    \end{subfigure}
    
    \begin{subfigure}[b]{0.8\linewidth}\centering

    \includegraphics[width=0.8\textwidth]{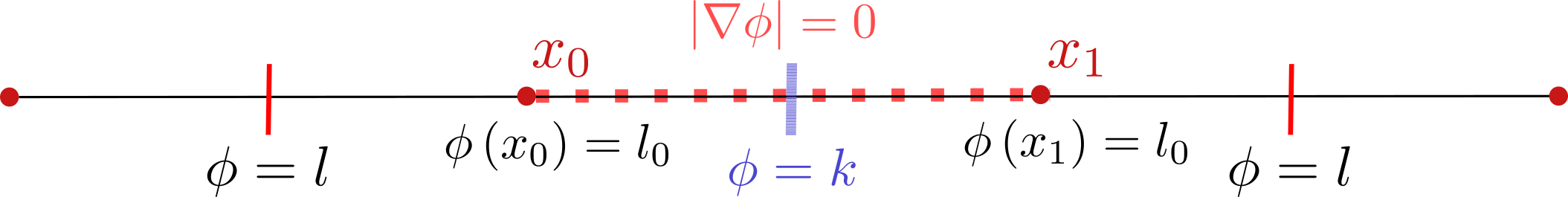}\label{1Dimension}
    \caption{}
    \end{subfigure}
    
    \caption{Example of pathological case during reinitialization procedure: \subref{1Dimension} for 1D; \subref{2Dimension} for 2D}
    \label{fig:illustration_gradnul_reinit}
\end{figure}

This method allows to focus the redistancing to elements close to the boundary. In most applications, this is sufficient as the signed distance function (SDF) property is only required in a neighborhood close to the $\Gamma$. However, for any application, in which interfaces merge, equidistant points will occur even in the narrow band. This needs to be considered in any redistancing scheme.  \\
In the next Section we detailed the finite element discretization of the proposed prediction-correction scheme.

\subsection{Finite element discretization}
In this section, we will first introduce the finite element discretization for a mesh that conforms to the interface $\Gamma$. Then, we will extend this formulation in case the mesh is not aligned with $\Gamma$.  \\

\subsubsection{Interface-fitted mesh}
Let $D_h$ denote the discretized domain $D$ into a mesh conforming to $\Gamma_h$. Using the space $V_h^0$ previously defined by \eqref{spaceVh_0}, the finite element problem to solve reads as follows.
\paragraph{Predictor}
Find $\phi_{p,h}\in V_h^{0}$ such that for all $v_h \in V_h^{0}$
\begin{equation}\label{eq:weak_predictor}
\int_{D_h}\nabla \phi_{p,h}\cdot\nabla v_h\,\text{ }dx=\int_{D_h}\sgn(x) v_h\text{ }dx + \int_{\partial D_{h}}\sgn(x)v_h\text{ }ds.
\end{equation}

\paragraph{Corrector}
With $\phi^0_h := \phi_{p,h}$ the initial value of the iterative scheme, the finite element element formulation to solve corrector problem becomes: For $n = 0 , ... ,N-1$, find $\phi_h^{n+1} \in V_h^0$ such that for all $v_h \in V_h^0$ 
\begin{equation}
    \int_{D_h} \nabla \phi_h^{n+1} \nabla v_h \, dx = \int_{D_h} \frac{\nabla \phi_h^n}{\left|\nabla\phi_h^n\right|} \nabla v_h \, dx . 
\end{equation}

We now extend this formulation to a finite element formulation for meshes that do not conform to the interface $\Gamma$. This choice is motivated by the fact that re-distancing is often applied in the context of moving boundary problems in which re-meshing can become challenging and costly. 

\subsubsection{Unfitted finite element formulation}
Let us now presume that the faces of the triangular or tetrahedral mesh $\mathcal{K}_{\;h}$ don't coincide with $\Gamma_h$. A new finite element space is constructed, which does not incorporate the boundary conditions on the interface. Within the continuous Galerkin framework, this space is defined as
\begin{equation}\label{eq:17}
V_{h}:=\left\{ v\in H^1\left(D\right)\cap C^{0}\left(D\right) \mid v_{\mid K}\in\left[\mathbb{P}_{1}\left(K\right) \text{ for all }K\in\mathcal{K}_{\;h}\right]\right\},
\end{equation}
where $K$ denotes the elements of the mesh. \\ 
Now that the Dirichlet boundary conditions are no longer included in the finite element space, we enforce them weakly through additional terms in our finite element formulation.  
\paragraph{Weak enforcement of boundary conditions}\label{BoundaryConstraint}
Three main methods exist to enforce boundary conditions weakly: i) the penalty method, ii) Langrange multiplier, iii) Nitsche's method. \\
In \cite{reinit_Basting}, the penalty method is employed. While it is relatively straightforward to formulate, it requires careful selection of a penalty parameter suitable for the numerical problem. If the parameter is chosen too low, the solution to the problem is not unique. Although the algorithm 
converges well to a reinitialized level set, insufficient penalization on the interface can lead to significant interface displacement. Conversely, choosing a penalty parameter that is too high results in excessive constraints, generating matrix ill-conditioning and subsequent boundary oscillations. This phenomenon is known as locking. \\
In \cite{AdamsEllipticPDEReinitDG}, the authors utilize the Lagrangian method to impose Dirichlet 
conditions on $\Gamma_h$. This method introduces a new unknown parameter, $\lambda$. At each iteration, the Lagrange multiplier $\lambda$ is updated by solving a PDE for the constraint. The choice of the Lagrange multiplier space is not always straightforward as it has to fulfill inf-sup stability conditions. \\
In this work, Nitsche's method \cite{Nitsche1971berEV} is used. Similar to the penalty method, it requires the selection 
of a penalty parameter. The choice of this parameter leads to difficulties akin to those previously mentioned
for the penalization method. However, unlike the penalty method, Nitsche's method offers a stable and effective numerical formulation for moderate penalty parameters. 

To integrate over $\Gamma_h$, which now intersects elements in an arbitrary fashion, we use the cut integration scheme introduced in the following section. \ref{CutIntegration}. 

\begin{figure}[H]
     \centering
     \includegraphics[width=0.6\textwidth]{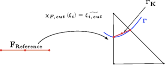}
     \caption{Illustration of mapping corresponding to sub-integration over $\Gamma_K$.}
     \label{fig:mapping_integration_facet}
 \end{figure}

\paragraph{Sub-integration for interface integral} \label{CutIntegration}
First, we reconstruct the interface in order to integrate over $\Gamma$ crossing element $K$. A simple linear interpolation of level set values is used in this work to calculate the points of intersection between the edges of the element $K$ and the interface and a piecewise linear interface, $\Gamma_K$, is reconstructed by connecting these points on the edges (see Figure~(\ref{fig:mapping_integration_facet})). 
Then, to integrate over the reconstructed interface part $\Gamma_K$, we generate a quadrature rule by mapping standard integration rules onto the piecewise linear approximation of $\Gamma_K$ inside the reference element. This generates a new quadrature rule with new quadrature points $\tilde{p}_{i}$ and new weights $\tilde{w}_{i}$ for each interface part. The mapping to generate the quadrature rule is mixed dimensional. An example of the approximation of the $\Gamma_K$ integral as part of $K$ is given in Figure~(\ref{fig:mapping_integration_facet}).

We can now formulate our predictor-corrector scheme for interfaces $\Gamma_h$ that freely intersect through a background mesh. 
\paragraph{Predictor}\label{h_Predictor}
Using the Nitsche method, the discretized weak formulation of the predictor problem reads: Find $\phi_{p,h} \in V_h$, such that for all $v_h \in V_h$ 
\begin{equation}\label{eq:weak_predictor_EF}
a\left(\phi_{p,h},v_h\right)=l\left(v_h\right),
\end{equation}
where
\begin{align}\label{eq:predictor}
a\left(\phi_{p,h},v_h\right) &= \int_{D_{h}}\nabla \phi_{p,h}\cdot\nabla v_h\,\text{ }dx -\int_{\Gamma_{h}}\nabla \phi_{p,h}\cdot \bfn_{\phi} \, v_h\text{ }ds-\int_{\Gamma_{h}}\nabla v_h\cdot \bfn_{\phi} \, \phi_{p,h}\text{ }ds \notag \\ 
& +\gamma_{D}\int_{\Gamma_{h}}h^{-1} \phi_{p,h} \, v_h\text{ }ds \\
l\left(v_h\right) &= \int_{D_{h}}\sgn(x)\cdot v_h\text{ }dx + \int_{\partial D_{h}}\sgn(x)v_h\text{ }ds.
\end{align}
The integration term over the discretization of $\Gamma$, denoted $\Gamma_h$, is done by the cut integration scheme introduce in the previous sub-Section \ref{CutIntegration}. The rest of the terms are integrated with a standard integration scheme. A cut integration on the volume is not required as the solution to our elliptic problem is continuous at $\Gamma$ despite the discontinuous RHS term $\sgn(x)$. 

\paragraph{Corrector}\label{h_Corrector}
Using Nitsche's method, and $\phi^0_h := \phi_{p,h}$ as the initial value of the iterative scheme, the finite element element formulation to solve corrector problem reads: For $n = 0 , ... ,N-1$, find $\phi_h^{n+1} \in V_h$ such that for all $v_h \in V_h$ 
\begin{equation}\label{eq:13}
a\left(\phi^{n+1}_{h},v_{h}\right)=l\left(v_h,\phi^{n}_{h}\right),
\end{equation}
where
\begin{align}
a\left(\phi^{n+1}_{h},v_{h}\right)&=\int_{D_{h}}\nabla\phi^{n+1}_{h}\cdot\nabla v_{h}\text{ }dx-\int_{\Gamma_{h}}\nabla\phi^{n+1}_{h}\cdot \bfn_{\phi} v_{h}\text{ }ds-\int_{\Gamma_{h}}\nabla v_{h}\cdot \bfn_{\phi}\phi^{n+1}_{h}\text{ }ds \notag \\
+&\gamma_{D}\int_{\Gamma_{h}} h^{-1}\phi^{n+1}_{h}v_{h}\text{ }ds\label{corrector_bilin}\\
l\left(v_{h},\phi^{n}_{h}\right)&=\int_{D_{h}}\frac{\nabla\phi^{n}_{h}}{\max\left(\left|\nabla\phi^{n}_{h}\right|,\epsilon\right)}\cdot\nabla v_{h}\text{ }dx \label{corrector_lin}
\end{align}
with $\epsilon$ being a very small positive scalar.

 
\section{Numerical Results}\label{Numerical_study}
In this section, we evaluate our prediction-correction method for four functions with increasing numerical complexity: (i) a smooth circular interface function, which enables a thorough assessment of the method (including convergence order, influence of the predictor problem, and analysis of the Nitsche parameter); (ii) a piecewise constant function; (iii) a complex example involving a sharply varying interface; and (iv) a three-dimensional case, demonstrating the ease with which our method can be extended to 3D problems.
To analyze the convergence towards the exact solution, we denote the difference between the numerical approximation, $\phi_{h}$, and the exact solution, $\phi_{ex}$ , as
\begin{equation}
e(\phi_{h}(x))=\phi_{h}(x) - \phi_{ex}(x), \quad x \in D_h.
\end{equation} 
We evaluate this error in the $L^{2}$ norm
\begin{equation}
\left\Vert e\right\Vert _{L^{2}\left(D_h\right)}^{2}=\frac{\int_{D_h}e^{2}(\phi_{h}(x))\,dx}{\int_{D_h}\,dx}.
\end{equation} 
\par
Furthermore, to evaluate the method's ability to converge towards the solution of the Eikonal equation \eqref{eq:signed_distance}, we introduce the Eikonal error, defined as
\begin{equation}
    e_{\text{eik}}(\phi_{h}(x))=1-\left|\nabla\phi_{h}(x)\right|, \quad x \in D_h.
\end{equation} 
We compute this error using the $L^{2}$ norm
\begin{equation}
    \left\Vert e_{\text{eik}}\right\Vert _{L^{2}\left(D_h\right)}^{2}=\frac{\int_{D}e^{2}_{\text{eik}}(\phi_{h}(x))\,dx}{\int_{D_h}\,dx}.
\end{equation} 
Finally to assess the method's precision in preserving the initial interface $\Gamma_h^0$ we define the 
following norm with respect to the  solution of the prediction problem $\phi_{h}^{0}$
\begin{align}\label{error_interface}
    \left\Vert \phi_{h}\right\Vert _{L^{2}{\left(\Gamma\right)}}^{2} &= \int_{\phi^{0}_{h}=0}\left(\phi_{h}(x)-\phi_{h}^{0}(x)\right)^{2}\text{ }ds \\
    &=\int_{\phi_{h}^{0}=0}\left(\phi_{h}(x)\right)^{2}\text{ }ds.
\end{align}

\subsection{Classical case: circular level set}\label{circle_func}
\begin{figure}[H]
\centering
    \begin{subfigure}[b]{0.29\linewidth}\centering
        \includegraphics[width=1\textwidth]{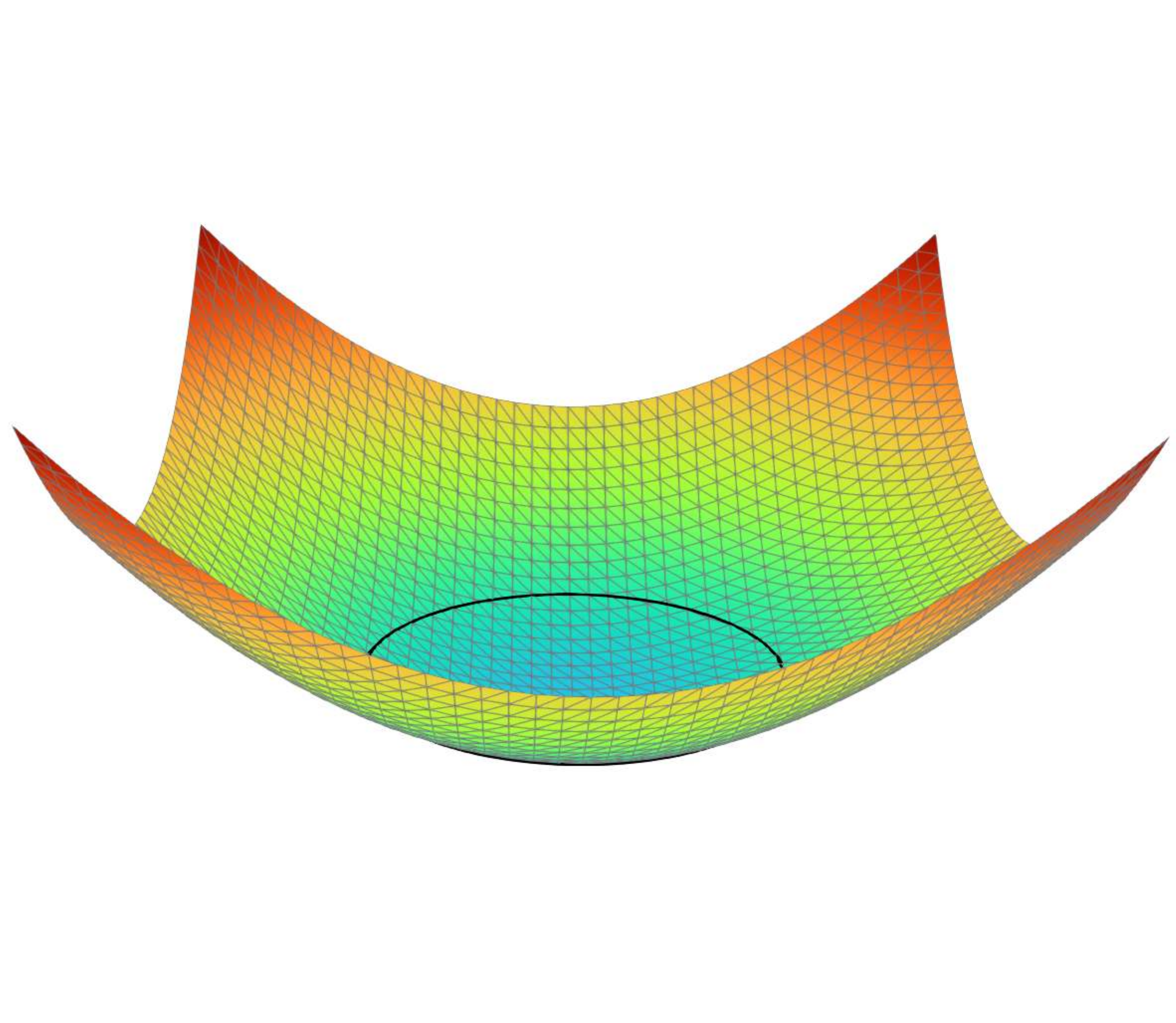}
        \caption{}\label{circle_init_PC_warp}
        
    \end{subfigure}
    \begin{subfigure}[b]{0.29\linewidth}\centering
         \includegraphics[width=1\textwidth]{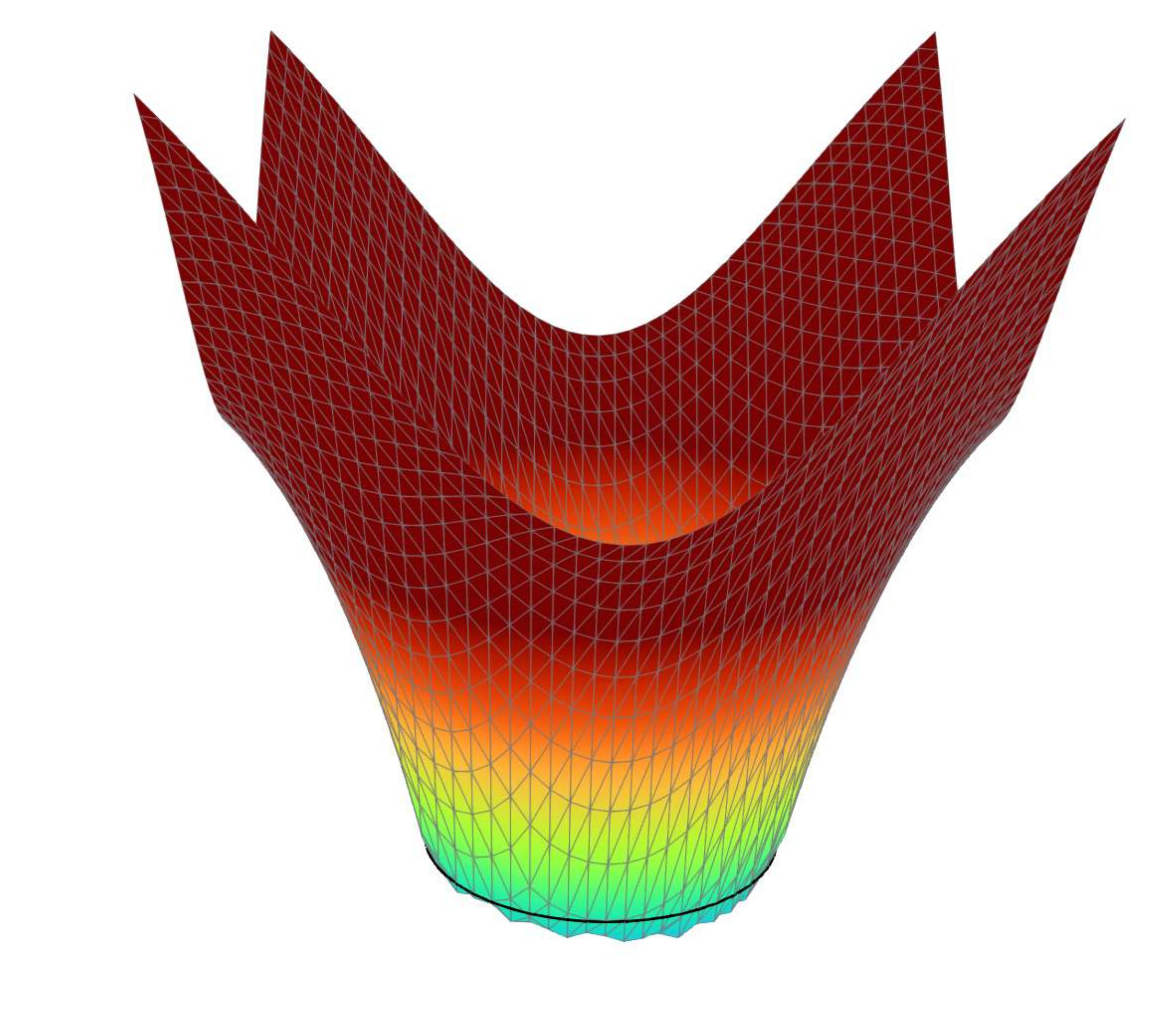}
         \caption{}
         \label{circle_predictor_PC_warp}
    \end{subfigure}
    
    \vspace{0.5cm}    
    \begin{subfigure}[b]{0.29\linewidth}\centering
         \includegraphics[width=1\textwidth]{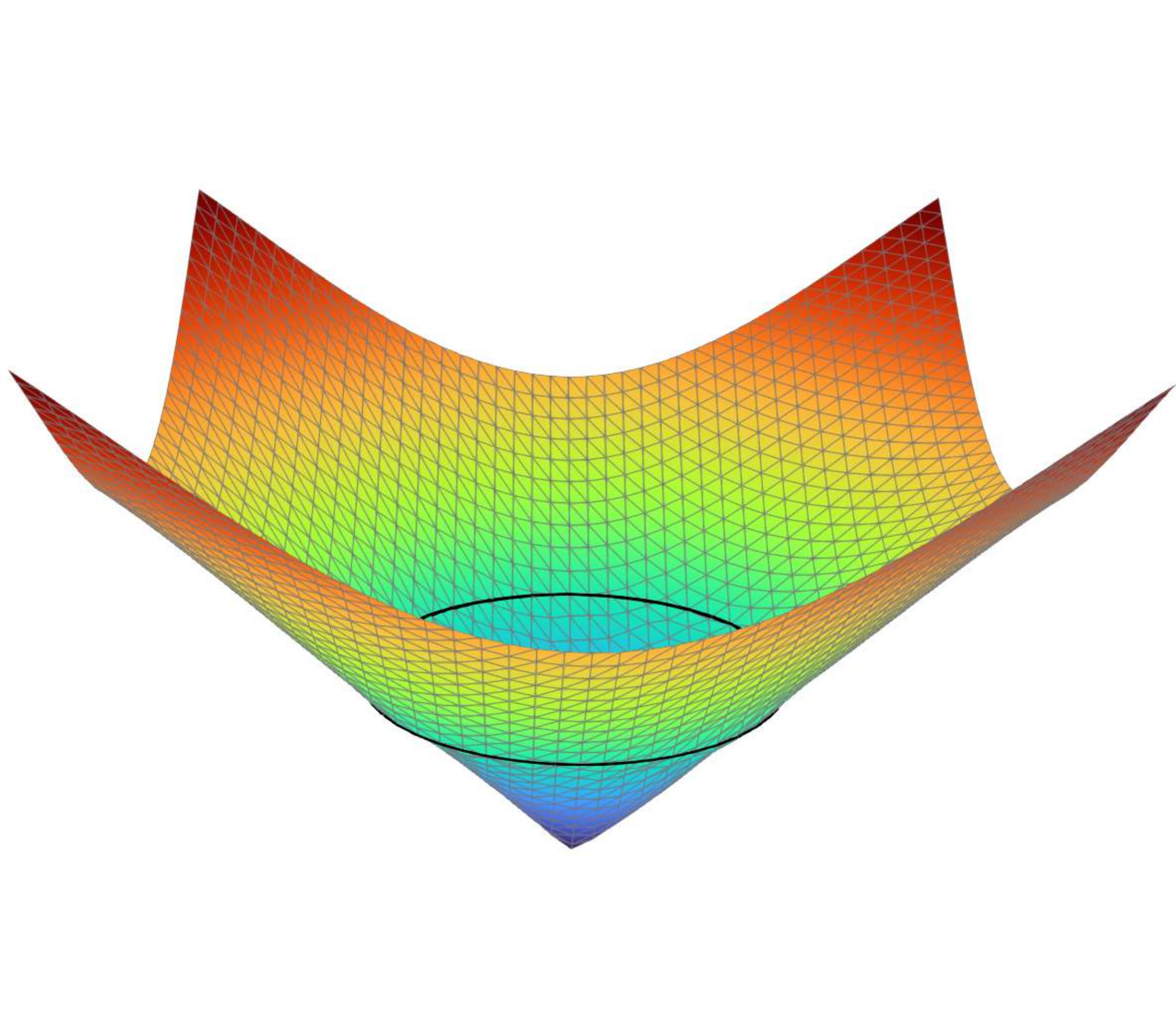}\centering
         \caption{}
         \label{circle_reinit_1PC_warp}
    \end{subfigure}
    \begin{subfigure}[b]{0.29\linewidth}\centering
         \includegraphics[width=1\textwidth]{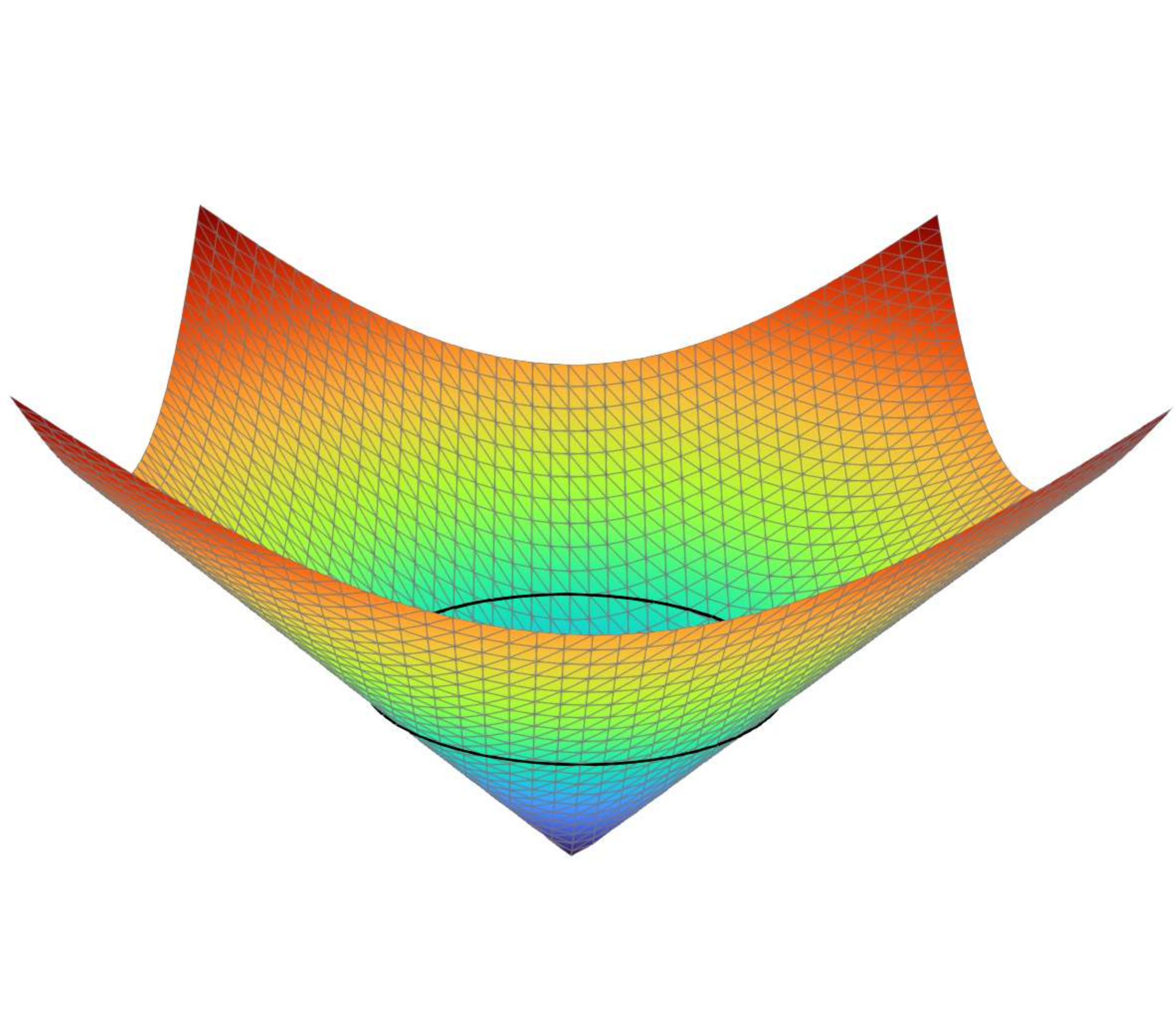}
         \caption{}
         \centering\label{circle_reinit_10PC_warp}
    \end{subfigure}
     \begin{subfigure}[b]{0.1\linewidth}\centering
         \includegraphics[width=0.9\textwidth]{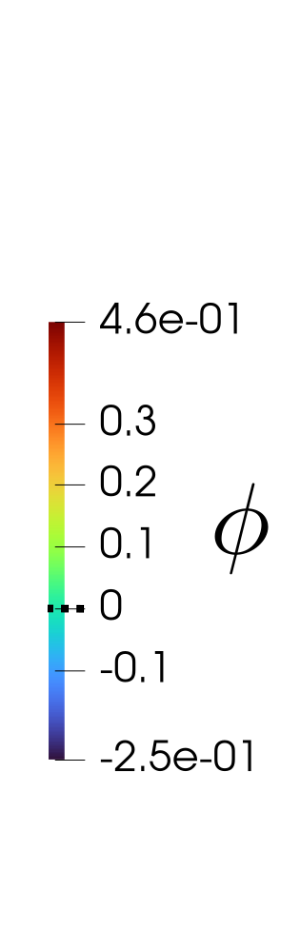}
    \end{subfigure}
    
    \caption{Circular level set for reinitialization with prediction-correction method with $h=\text{L}/40$ and $\gamma_{D}=10^{4}$: \subref{circle_init_PC_warp} level set warped before reinitialization; \subref{circle_predictor_PC_warp} level set warped after prediction problem; \subref{circle_reinit_1PC_warp} predicted level set warped after 1 iteration of correction; \subref{circle_reinit_10PC_warp} level set warped after 10 iterations of correction.}
    \label{fig:circle_warped}
\end{figure}

This first numerical example is used to study the order of convergence of the prediction-correction method for the three norms introduced above. We start with an initial quadratic function on the domain $D=\left[0,\text{L}\right]^{2}$, with $\text{L} = 1$,  defined as
\begin{equation}\label{def_ls_circle}
    \phi_{\text{init}}\left(x,y\right)=\left(x-x_{0}\right)^{2}+\left(y-y_{0}\right)^{2}-r^{2}
\end{equation}
with $\left(x_{0},y_{0}\right)$ being the coordinate of the center of the circle and $r$ being the radius. We set $\left(x_{0},y_{0}\right)=\left(0.5,0.5\right)$ and $r=0.25$. As shown in Figure~(\ref{circle_init_PC_warp}), this function is a parabole, and not a signed distance function. 
The exact solution of the reinitialized level set, denoted $\phi_{ex}$, is known as a cone centered at $\left(x_{0},y_{0}\right)$ and such that the line of level zero coincides with that of $\phi_{\text{init}}$. The exact solution $\phi_{ex}$  is given by
\begin{equation}
    \phi_{ex}\left(x,y\right)=\sqrt{\left(x-0.5\right)^{2}+\left(y-0.5\right)^{2}}-r.
\end{equation}

\begin{figure}[H]
\centering
    \begin{subfigure}[b]{0.29\linewidth}\centering
        \includegraphics[width=0.9\textwidth]{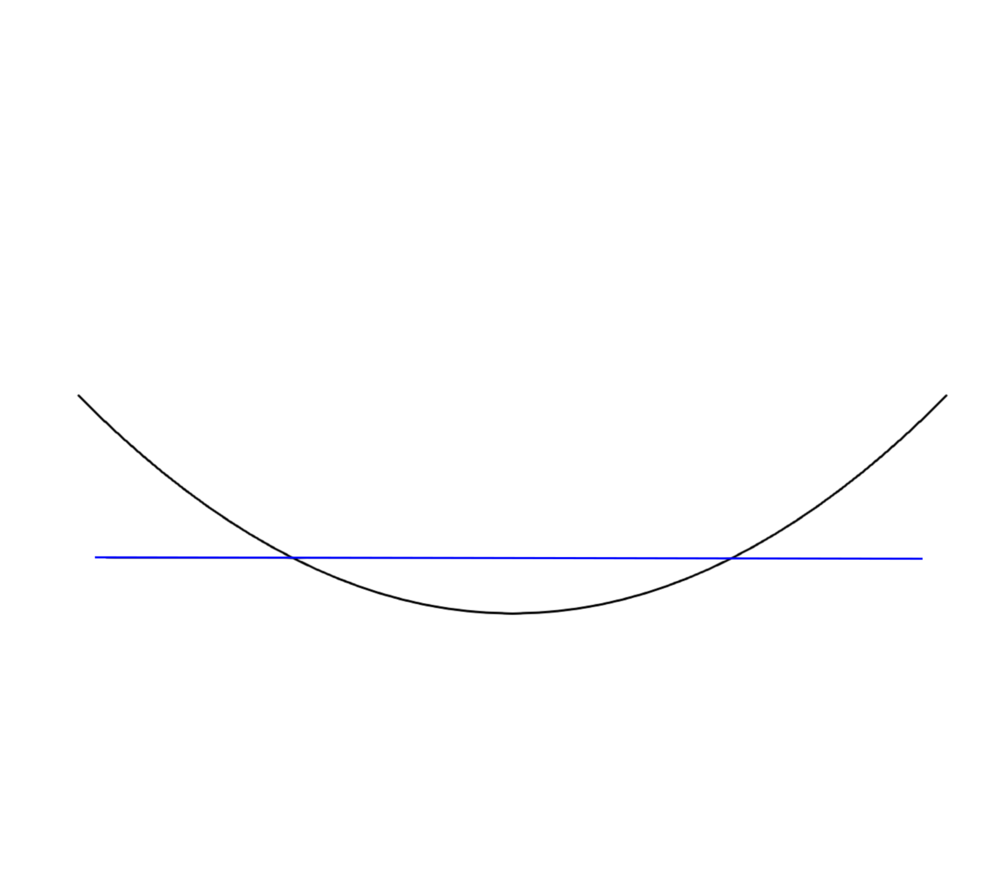}
        \caption{}\label{circle_line_init}
    \end{subfigure}
    \begin{subfigure}[b]{0.29\linewidth}\centering
         \includegraphics[width=0.9\textwidth]{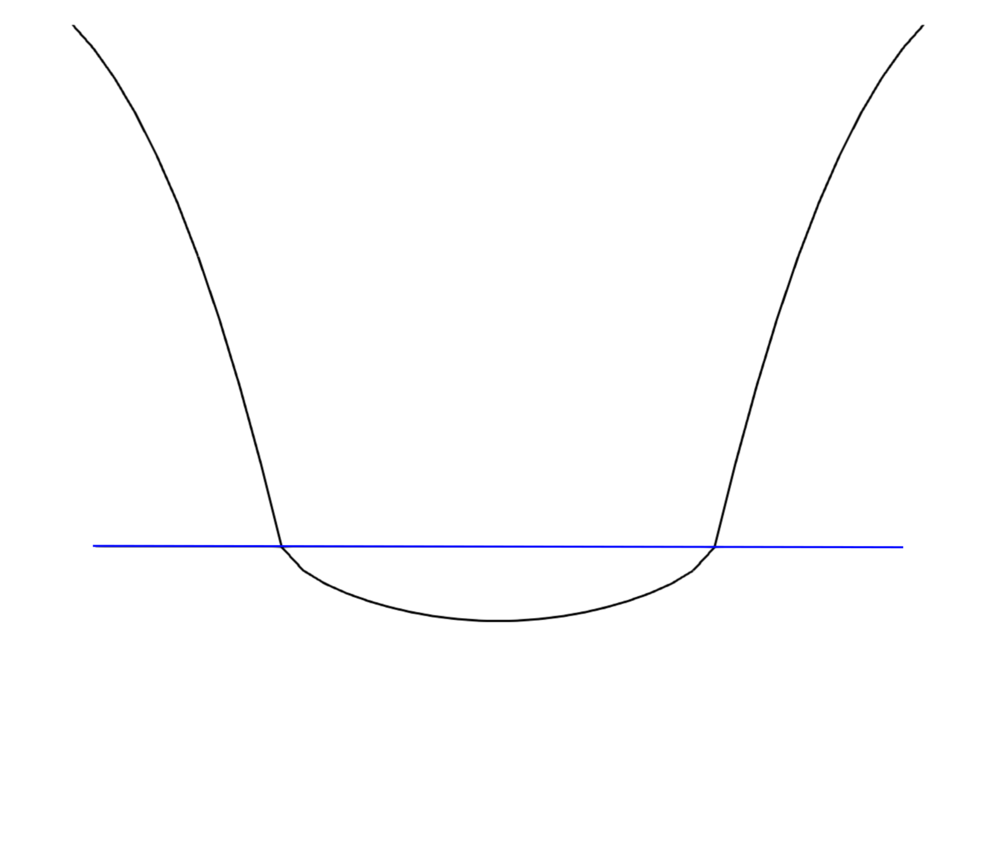}
        \caption{}\label{circle_line_predictor}
    \end{subfigure}
    \begin{subfigure}[b]{0.29\linewidth}\centering
         \includegraphics[width=0.9\textwidth]{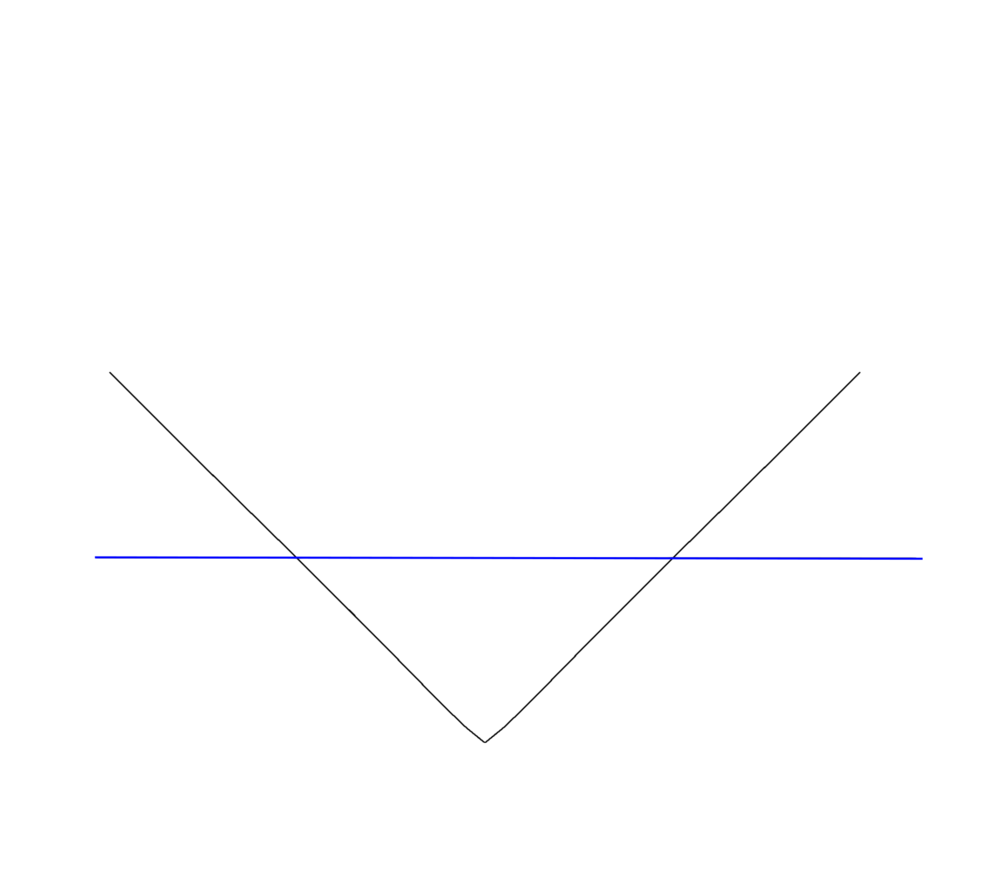}
        \caption{}\label{circle_line_corrector10ite}
    \end{subfigure}

    \caption{View of level set function along centre line: \subref{circle_line_init} initial function; \subref{circle_line_predictor} solution of predictor step; \subref{circle_line_corrector10ite} solution of correction step (after 10 iterations).}
    \label{fig:circle_centre_line}
\end{figure}

\begin{figure}[H]
\centering
 \begin{subfigure}[b]{0.45\linewidth}\centering
         \includegraphics[width=1\textwidth]{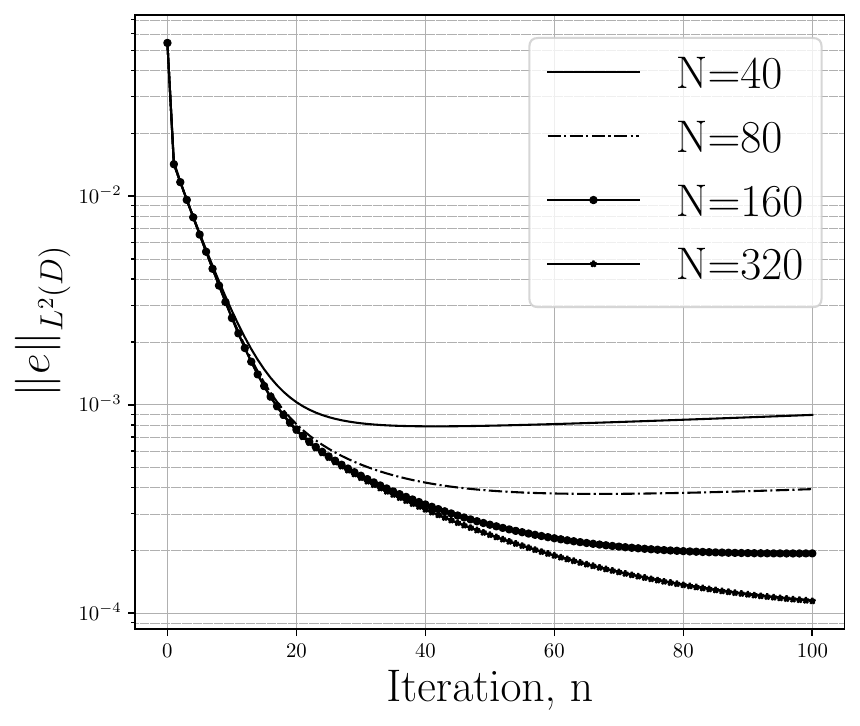}
\caption{}\label{cv_error_e}
    \end{subfigure}
    \begin{subfigure}[b]{0.45\linewidth}\centering
        \includegraphics[width=1\textwidth]{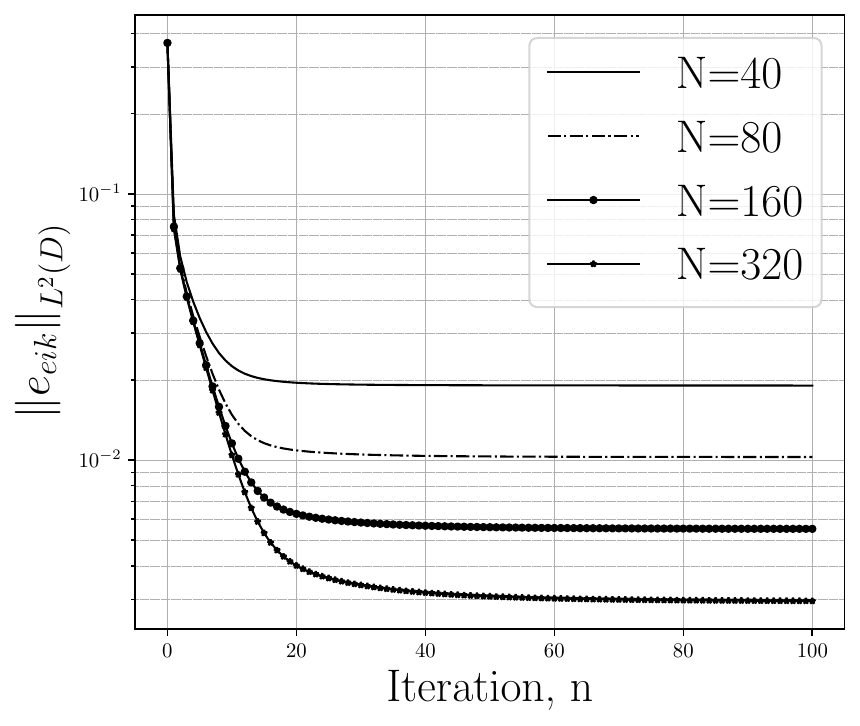}
        \caption{}\label{cv_error_eik} 
    \end{subfigure}
    \caption{Error evolution of circle example (prediction error is not plotted) with $\gamma_{D}=10^{4}$ for different mesh sizes: \subref{cv_error_e} Error $\left\Vert e\right\Vert_{L^{2}\left(D\right)}$ ; \subref{cv_error_eik} Error $\left\Vert e_{\text{eik}}\right\Vert_{L^{2}\left(D\right)}$.}
    \label{level_circle_ls_error}
\end{figure}

\paragraph{Convergence study}
We fix the Nitsche's parameter to $10^{4}$ and we study the errors defined above for different mesh sizes.
By observing the evolution of the $L^2$-errors of $e$ and $e_{\text{eik}}$ as a function of iterations 
(see Figures (\ref{cv_error_e}) and (\ref{cv_error_eik})), we note that, regardless of the mesh size, these errors fall below 0.01 and 0.05, respectively, after just 3 iterations. The level set function after one iteration of the correction problem (see Figure (\ref{circle_reinit_1PC_warp})) and after ten iterations (see (\ref{circle_reinit_10PC_warp})) differ only very slightly.
In practice, this means that only one iteration of the correction problem gives a sufficiently accurate approximation to the Eikonal equation for any application in which a strict verification of the Eikonal equation is not required, and an approximation within these error margins is acceptable (for example, in shape optimization with the level set method). 
The cross-section displayed in Figure~(\ref{fig:circle_centre_line}) shows that even at the center of the domain, near the singularity, the level set shows no oscillation or locking effects.

\paragraph{Convergence order}
In this section, we calculate the convergence order of the prediction-correction scheme for the three errors defined in the beginning of this section. We set $\gamma_D = 10^4$ for the convergence study. The solution is considered to have converged if 

\begin{equation}
   \left| \left\Vert e_{\text{eik}}(\phi^{n+1}) \right\Vert _{L^{2}\left(D\right)} - \left\Vert e_{\text{eik}}(\phi^{n}) \right\Vert _{L^{2}\left(D\right)} \right|<10^{-8}.
\end{equation}

We study convergence with mesh refinement for the series of triangular meshes with size $h=\text{L}/60,\text{L}/80,\text{L}/120,\text{L}/160$.

\begin{figure}[H]
\centering
    \begin{subfigure}[b]{0.45\linewidth}\centering
        \includegraphics[width=1\textwidth]{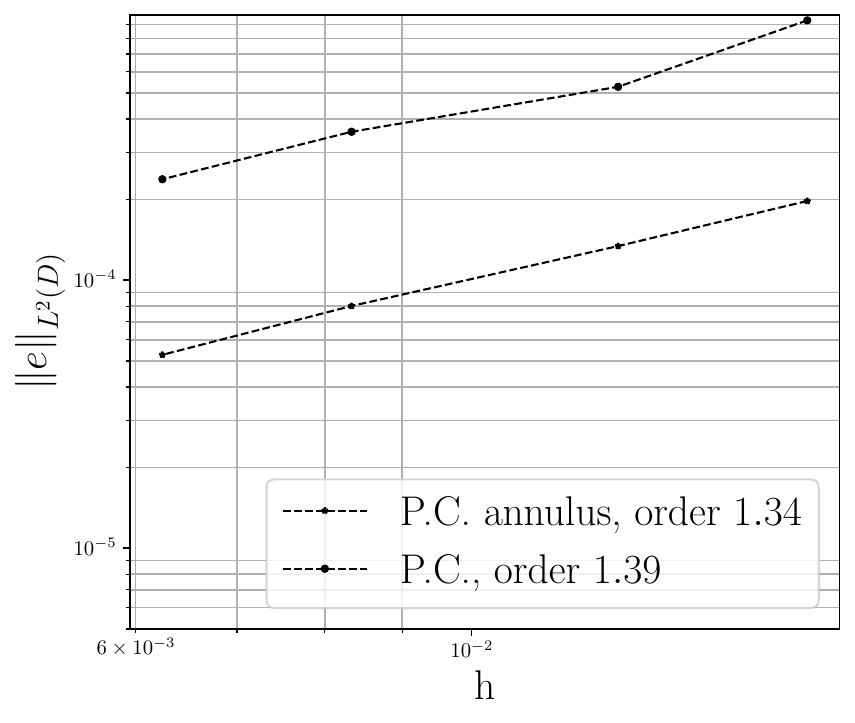}
        \caption{}\label{order_interface_circle}
    \end{subfigure}
    \begin{subfigure}[b]{0.45\linewidth}\centering
         \includegraphics[width=1\textwidth]{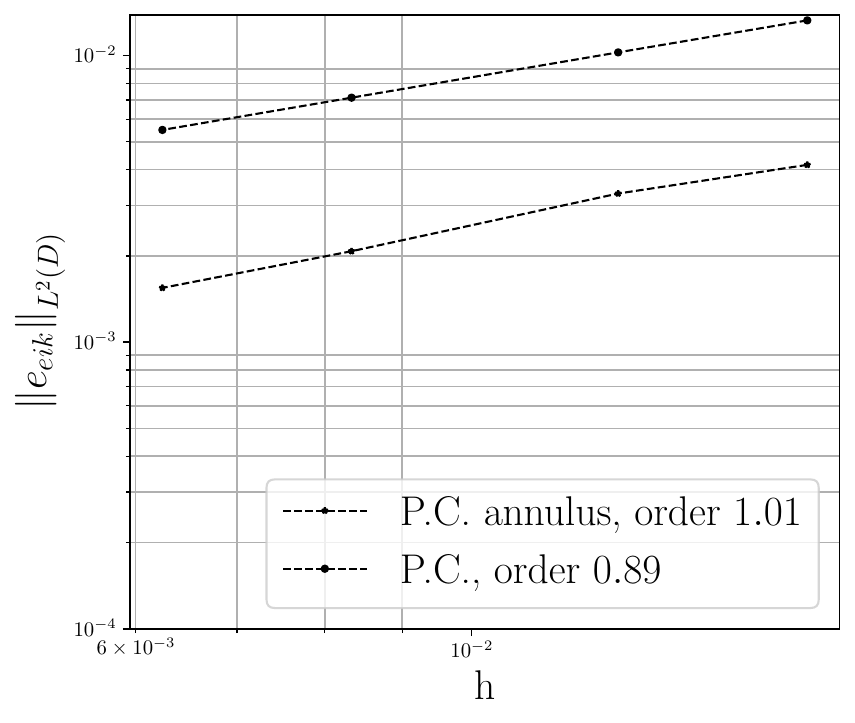}
        \caption{}\label{error_circle_order_eik}
    \end{subfigure}
    \vspace{0.5cm} 
        \begin{subfigure}[b]{0.45\linewidth}\centering
            \includegraphics[width=1\textwidth]{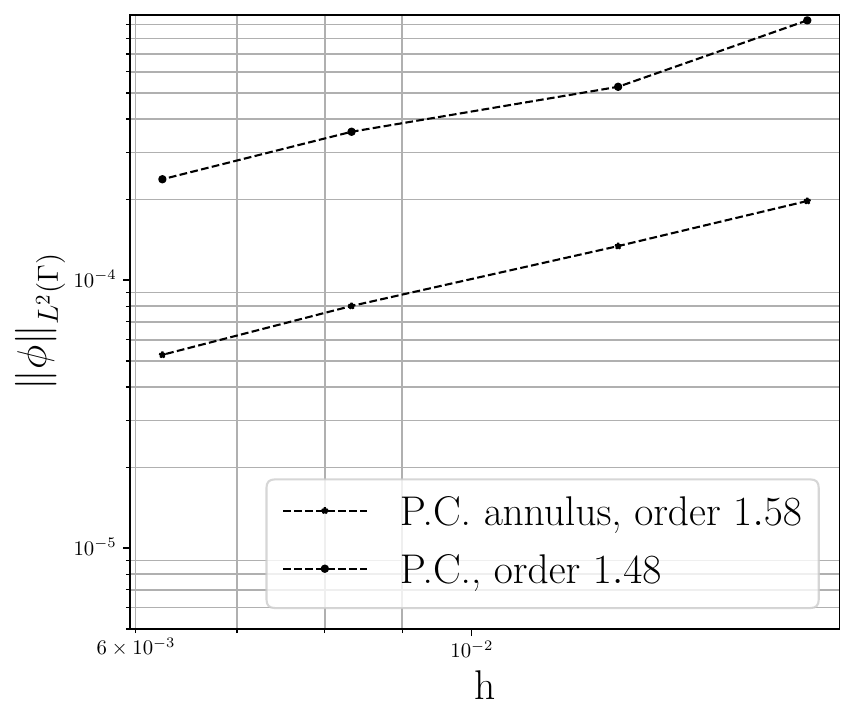}\\
            \caption{}
    \label{error_phi_circle}
        \end{subfigure}
    \caption{The convergence order of the P.C. method with respect to the following errors: \subref{order_interface_circle} Error $\left\Vert e\right\Vert_{L^{2}\left(D\right)}$ ; \subref{error_circle_order_eik} Error $\left\Vert e_{\text{eik}}\right\Vert_{L^{2}\left(D\right)}$ ; \subref{error_phi_circle} Error $ \left\Vert \phi \right\Vert_{L^{2}\left(\Gamma\right)}$.}
    \label{OrderPC}
\end{figure}

The order of convergence is computed between the errors obtained with $h=\text{L}/60$ and $h=\text{L}/160$.
Numerically, we observe in Figures~(\ref{order_interface_circle}), (\ref{error_circle_order_eik}) and (\ref{error_phi_circle}) 
that the convergence 
order of our scheme on the interface is $1.48$, the order of the Eikonal error is $0.89$ and the order for the error $e$ in the whole domain is $1.39$.  These convergence rates are below optimal which would be order 2 for the $L^2$-errors and 1 for the Eikonal error. We have two main sources of error: (i) equidistant points with norm of the gradient close to zero (theoretically not differentiable), (ii) the splitting error in the corrector scheme from the Picard fixed point linearization. To test if the point in the centre of the circle is causing the reduction in the convergence order, we solve the circular level set problem only in a narrow band around the interface. The errors are denoted as "P.C. annulus" in Figures~(\ref{order_interface_circle}), (\ref{error_circle_order_eik}) and (\ref{error_phi_circle}). We see that while the $L^2$ error does not significantly improve, we recover an optimal order of convergence for the Eikonal error. 

\paragraph{Comparison with Elliptic method:}
In this section, we compare our method to the Elliptic one. The elliptic method can be used for this example because the norm of the gradient is not close to zero in large areas. Table~\ref{tab:predictor_study} shows the errors at convergence, which remain unchanged regardless of whether the predictor problem is solved or not. This demonstrates that our method does not degrade the solution provided by conventional reinitialization approaches such as the elliptic based reinitialization.

The downside of using our method in such cases is that it requires a greater number of iterations to reach convergence compared to the classical minimization based reinitialization. In this example, the number of iterations increases by a factor of approximately 1.25. 

However, the elliptic method fails to converge if the initial level set function contains large flat areas of very small values for the norm of the gradient. In these cases, solving the predictor problem becomes essential. We will demonstrate this in the next section where we take an initial flat function. 
\begin{table}[H]\centering
\begin{tabular}{lrrrlrrr}
  \toprule
   & \multicolumn{1}{c}{Number of iteration, n} && $\left\Vert e_{\text{eik}}\right\Vert _{L^{2}\left(D\right)}$  && \multicolumn{1}{c}{$\left\Vert e\right\Vert _{L^{2}\left(D\right)}$} && \multicolumn{1}{c}{$\left\Vert e\right\Vert _{L^{2}\left(\Gamma\right)}$} \\
  \midrule
  Elliptic   & 279      &&  0.01026 && 0.00041 && 0.00027 \\
  Predictor-Corrector  & 351     &&   0.01026 &&  0.00042 && 0.00027\\
  \bottomrule
\end{tabular}
\caption{Comparison between our method and the minimization method, for $N=80$, $\gamma_{D}=10000$ for minimization scheme and $\gamma_{D}=20000$ for predictor-corrector scheme.}
\label{tab:predictor_study}
\end{table}

\subsection{Flat function}\label{flat_func}
\begin{figure}[!ht]
\centering
    \begin{subfigure}[b]{0.27\linewidth}\centering
        \includegraphics[width=0.9\textwidth]{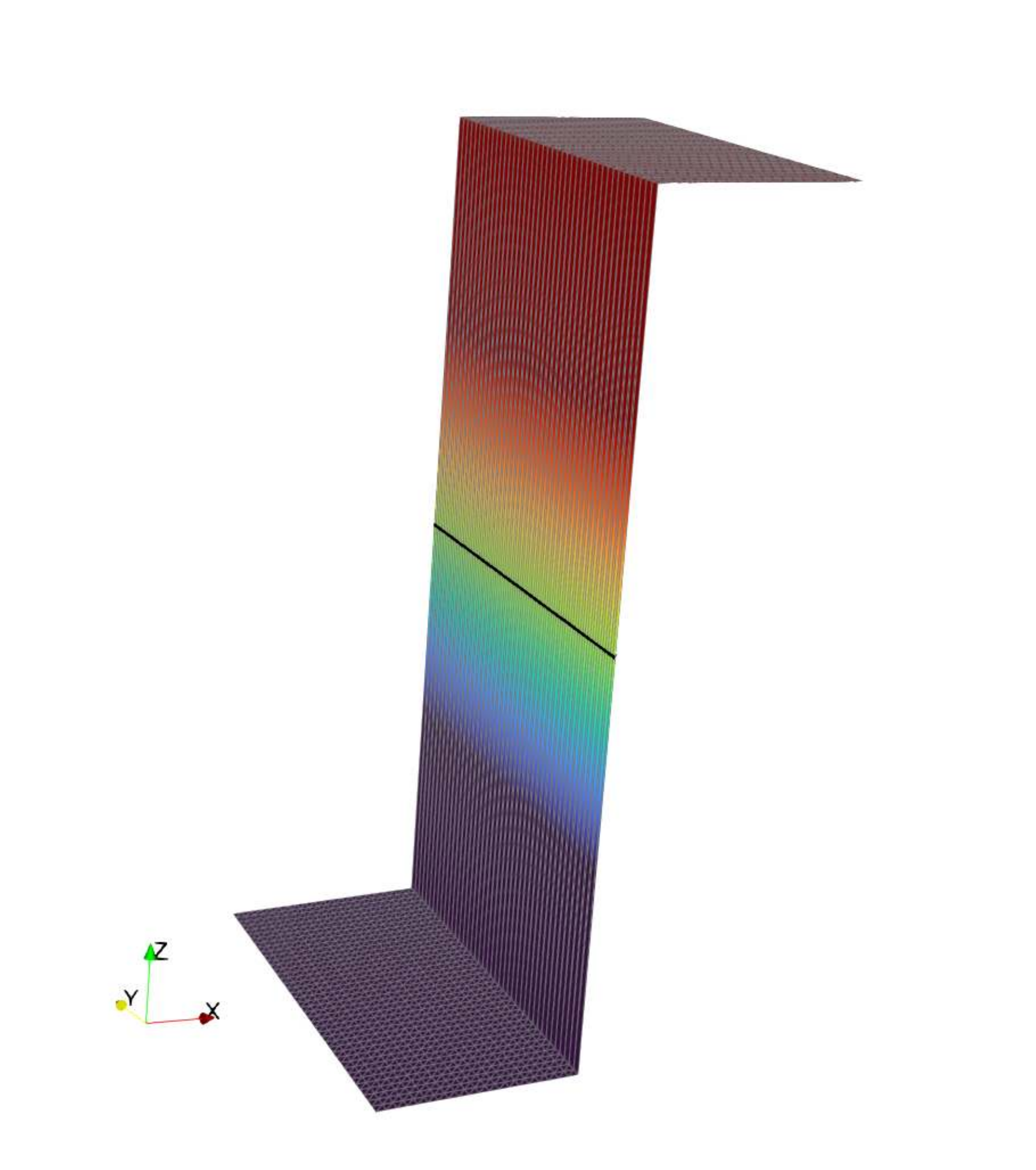}
        \caption{}\label{stepfun_init_PC_warp}
    \end{subfigure}
    \begin{subfigure}[b]{0.27\linewidth}\centering
         \includegraphics[width=0.9\textwidth]{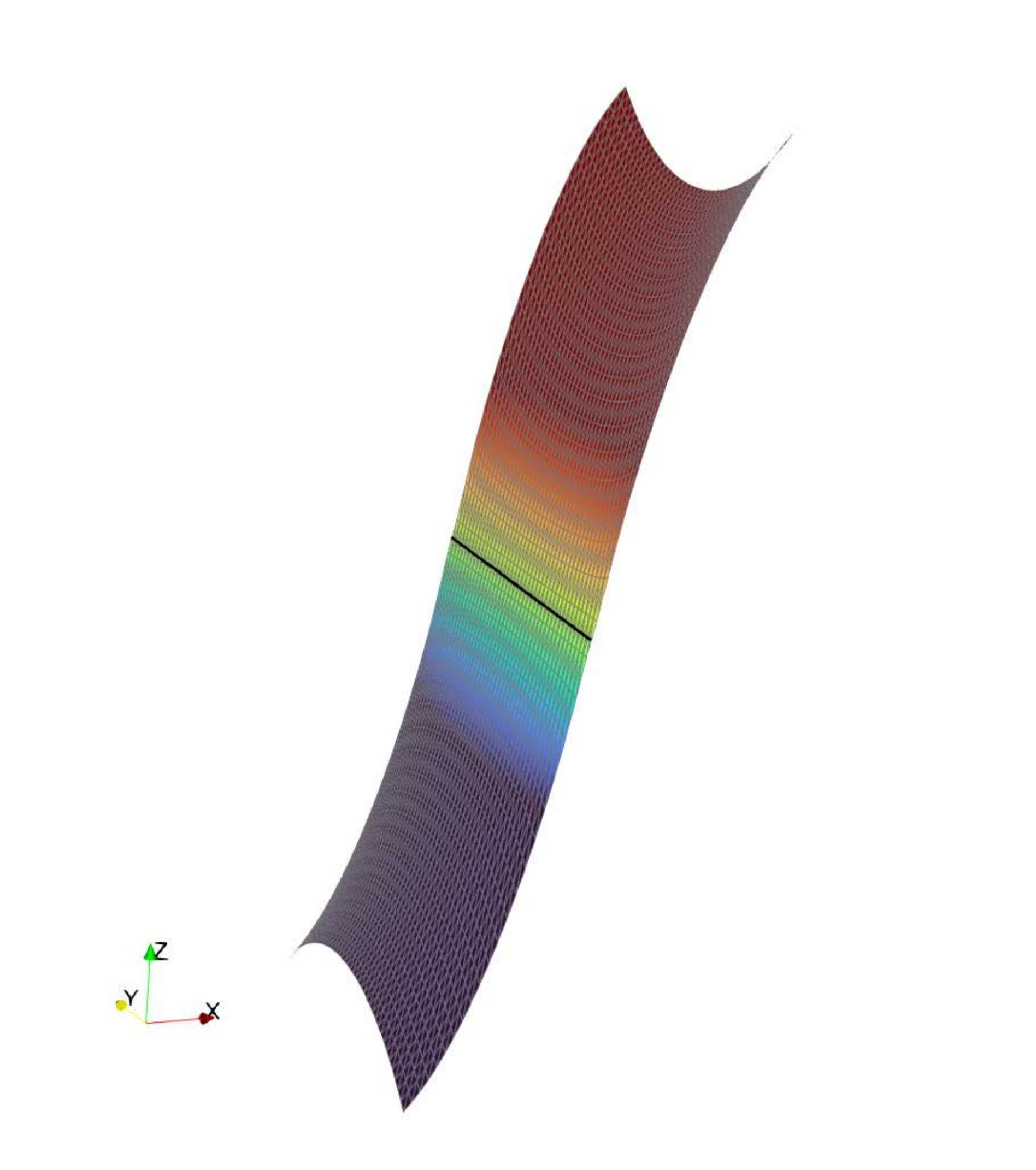}
        \caption{}\label{stepfun_corrector_PC_warp}
    \end{subfigure}
    \begin{subfigure}[b]{0.27\linewidth}\centering
         \includegraphics[width=0.9\textwidth]{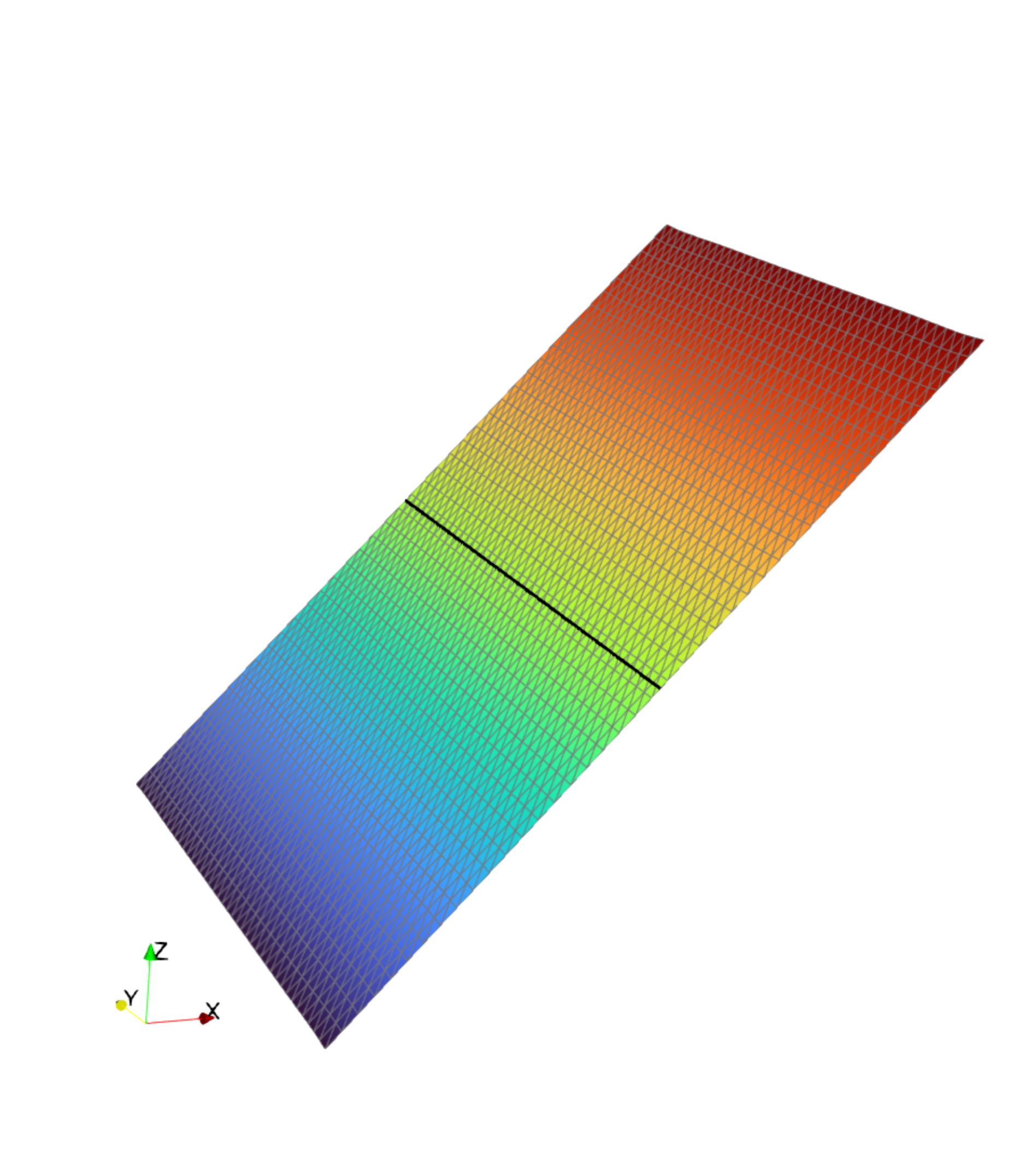}
        \caption{}\label{stepfun_reinit_PC_warp}
    \end{subfigure}
    \begin{subfigure}[b]{0.1\linewidth}\centering
         \includegraphics[width=0.9\textwidth]{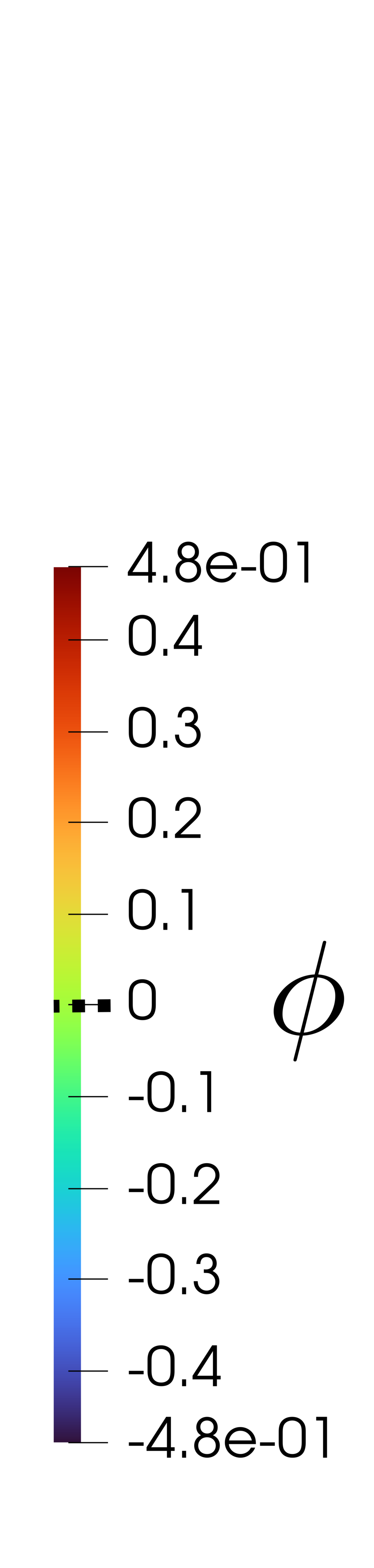}
    \end{subfigure}
    \caption{Step-shape warped for reinitialization step with P.C. method, $\gamma_{D}=10$ and $h=\text{L}/40$: \subref{stepfun_init_PC_warp} Level set function before reinitialization; \subref{stepfun_corrector_PC_warp} level set function after prediction problem; \subref{stepfun_reinit_PC_warp} level set function after P.C. reinitialization.}\label{fig:init_step_func}
\end{figure}

A second example is studied to understand the capability of the scheme to correct a flat function, i.e. large area with $|\nabla \phi| \approx 0$. These type of flat functions are a challenge for the minimization based reinitialization scheme which includes a $\frac{1}{|\nabla \phi|}$ term.  \\
Let the initial level set function be a step function (see  Figure~(\ref{fig:init_step_func})) on the domain $D=\left[0,\text{1}\right]^{2}$ defined by:
 
\begin{equation}
    \phi_{\text{init}}(x,y)=\begin{cases}
-1 & \text{if }x>0.5,\\
0 & \text{if }x=0.5,\\
1 & \text{if }x<0.5.
\end{cases}
\end{equation}

\begin{figure}[!ht]
\centering
    \begin{subfigure}[b]{0.27\linewidth}\centering
        \includegraphics[width=0.9\textwidth]{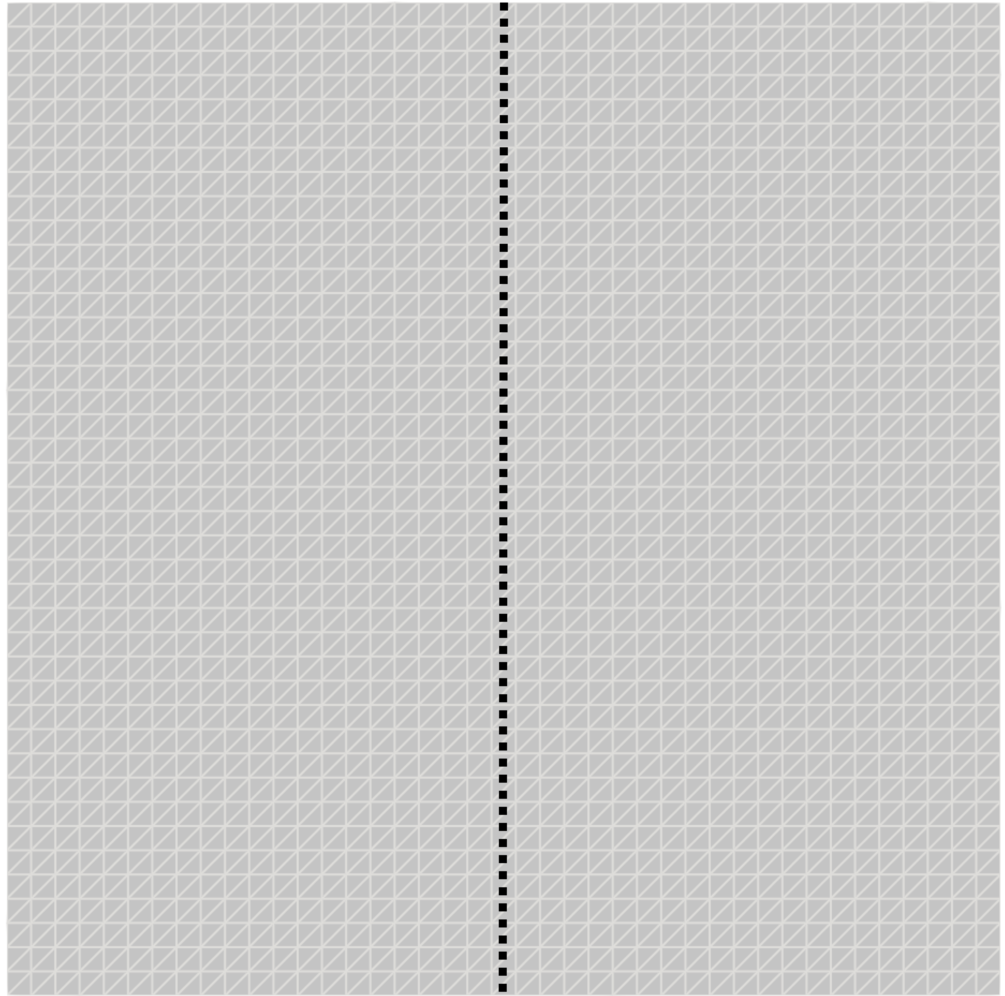}
        \caption{}\label{stepfun_init_PC_contour}
    \end{subfigure}
    \begin{subfigure}[b]{0.27\linewidth}\centering
         \includegraphics[width=0.9\textwidth]{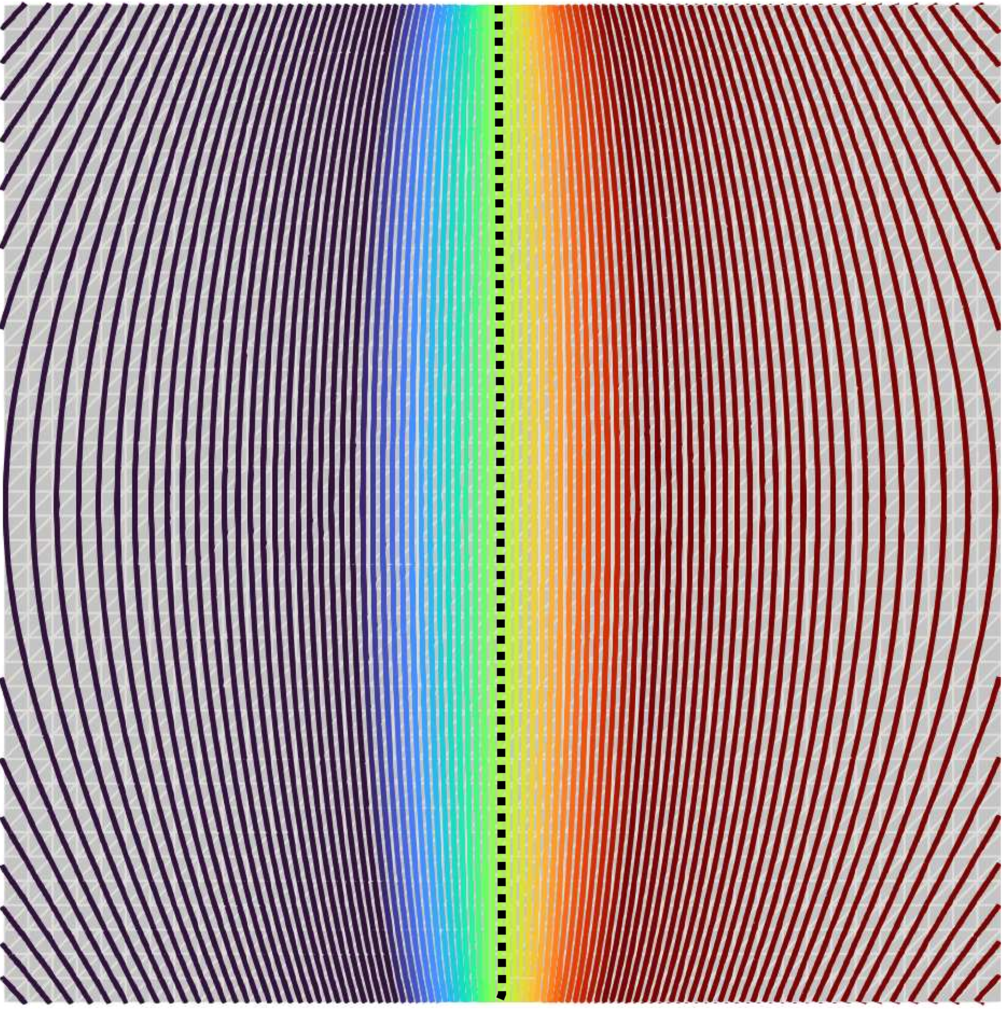}
         \caption{}\label{stepfun_corrector_PC_contour}
    \end{subfigure}
    \begin{subfigure}[b]{0.27\linewidth}\centering
         \includegraphics[width=0.9\textwidth]{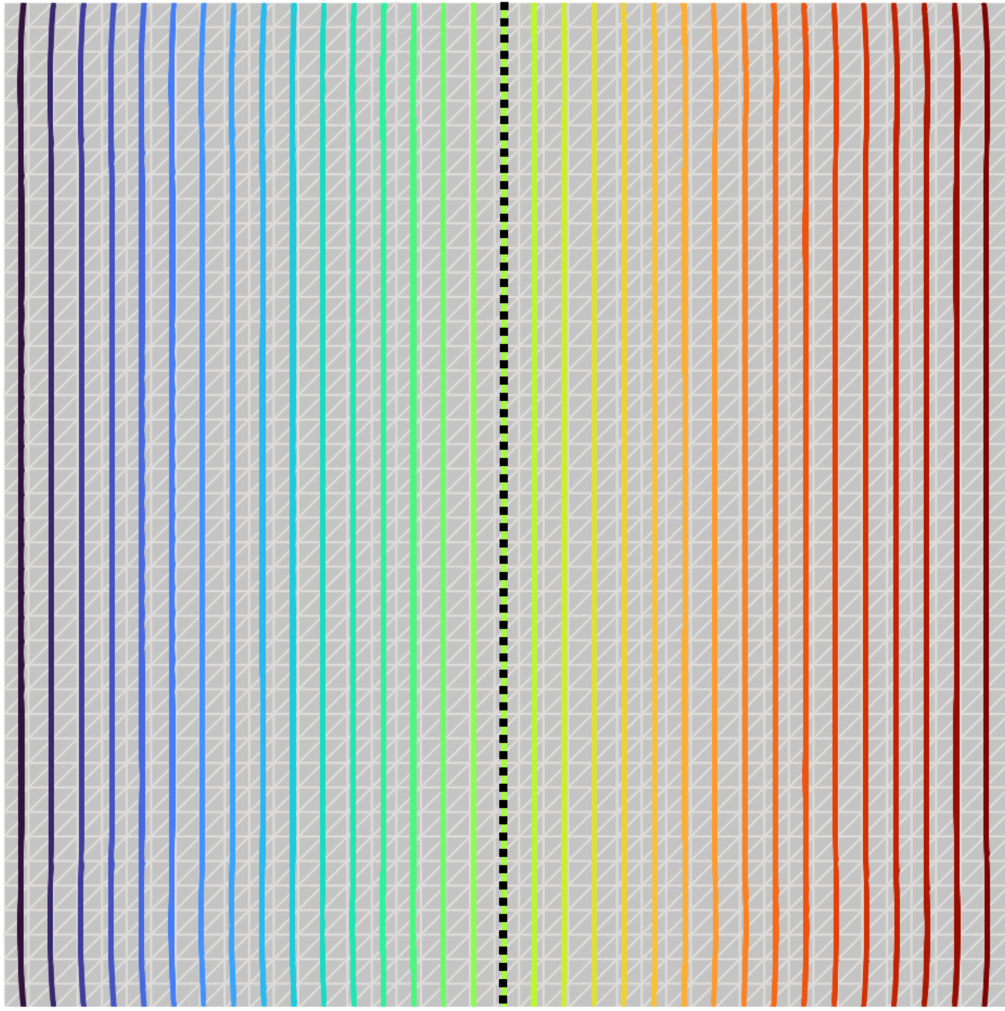}
         \caption{}\label{stepfun_reinit_PC_contour}
    \end{subfigure}
    \begin{subfigure}[b]{0.1\linewidth}\centering
         \includegraphics[width=0.9\textwidth]{legend_step_func.png}
    \end{subfigure}
    \caption{Step-shape level set contours for reinitialization step with P.C. method, $\gamma_{D}=10$ and $h=\text{L}/40$: \subref{stepfun_init_PC_contour} level set contours before reinitialization; \subref{stepfun_corrector_PC_contour} level set contours after predictor problem; \subref{stepfun_reinit_PC_contour} level set contours after P.C. reinitialization.}\label{level_set_step_ls}
\end{figure}

Without a predictor problem, the initial minimization problems right hand side becomes infinity everywhere in the Picard fixed point iteration scheme.
The solution of the predictor problem is a level set with the same sign as the initial level set, while correcting the zero gradient norms so that the corrector problem is well-defined. As observed in Figures (\ref{fig:init_step_func}) and (\ref{level_set_step_ls}) for the level set solution of the predictor problem, no element $K$ of the mesh satisfies: $| \nabla \phi(K) |=0$. From this predictor problem, only a few iterations are necessary to obtain a reinitialized level set function.

\begin{figure}[!ht]
\centering
    \begin{subfigure}[b]{0.27\linewidth}\centering
        \includegraphics[width=0.9\textwidth]{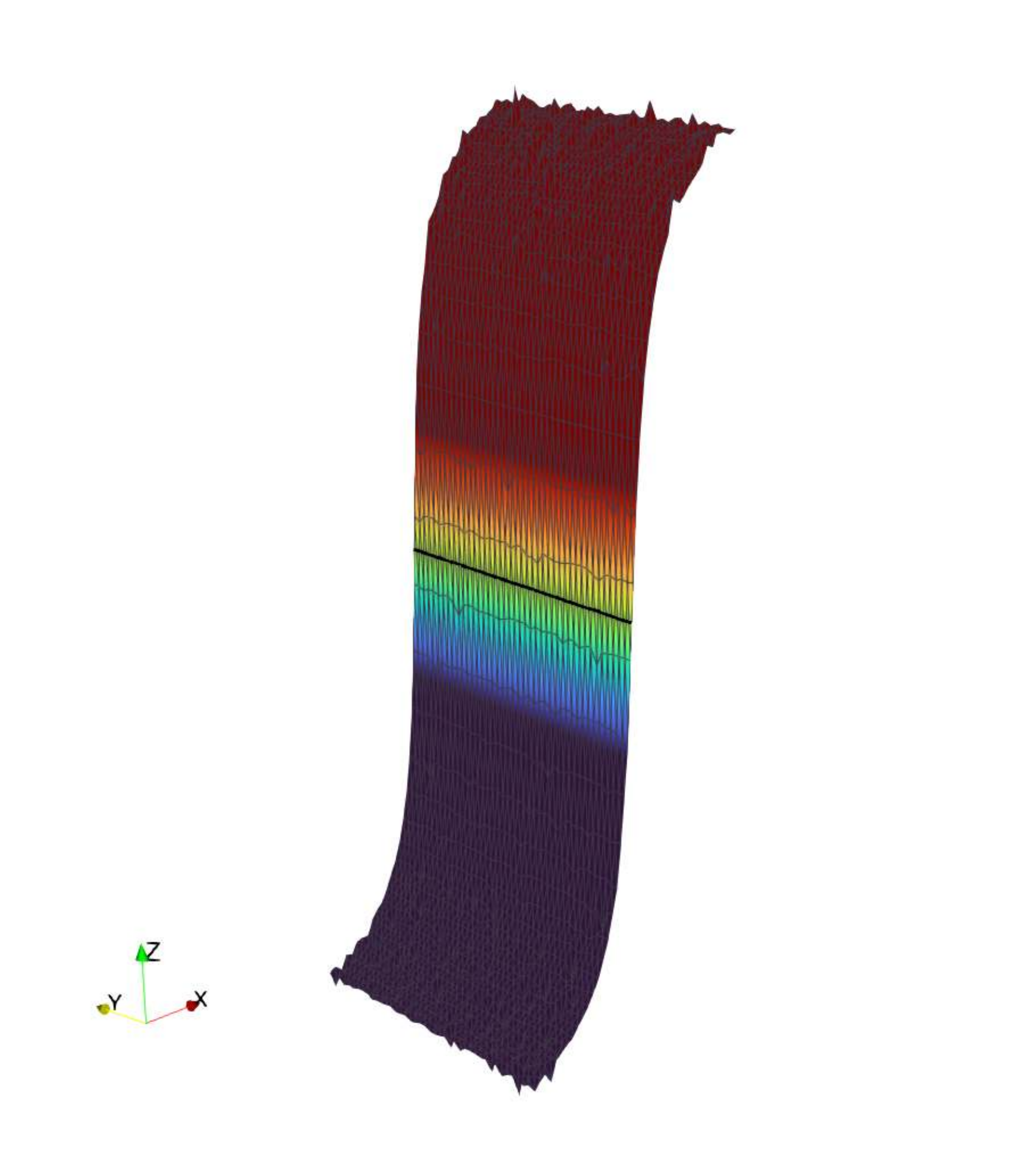}
        \caption{}\label{ls_init_PC_warp}
    \end{subfigure}
    \begin{subfigure}[b]{0.27\linewidth}\centering
         \includegraphics[width=1.\textwidth]{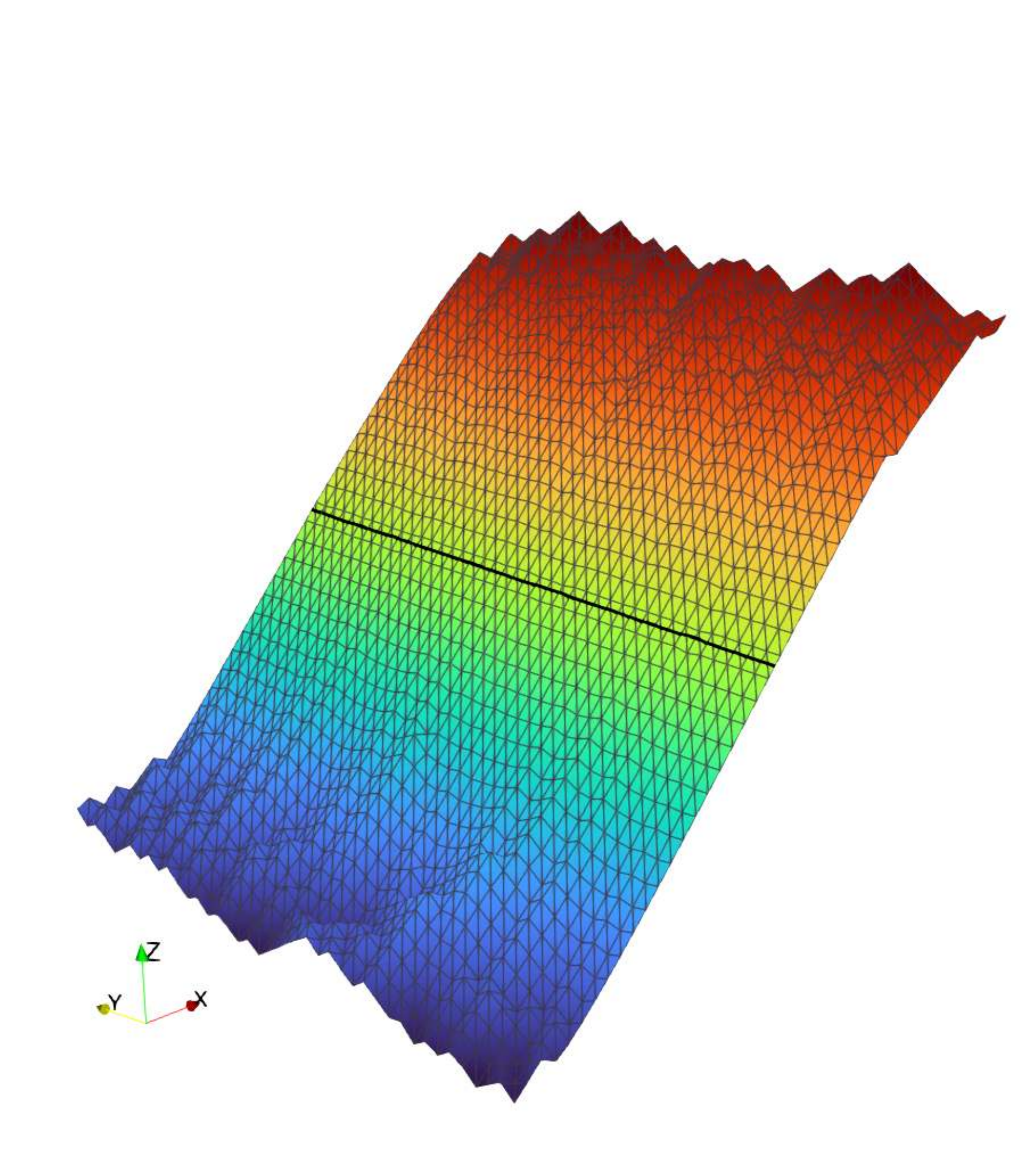}
        \caption{}\label{ls_elliptic_warp}
    \end{subfigure}
    \begin{subfigure}[b]{0.27\linewidth}\centering
         \includegraphics[width=1.\textwidth]{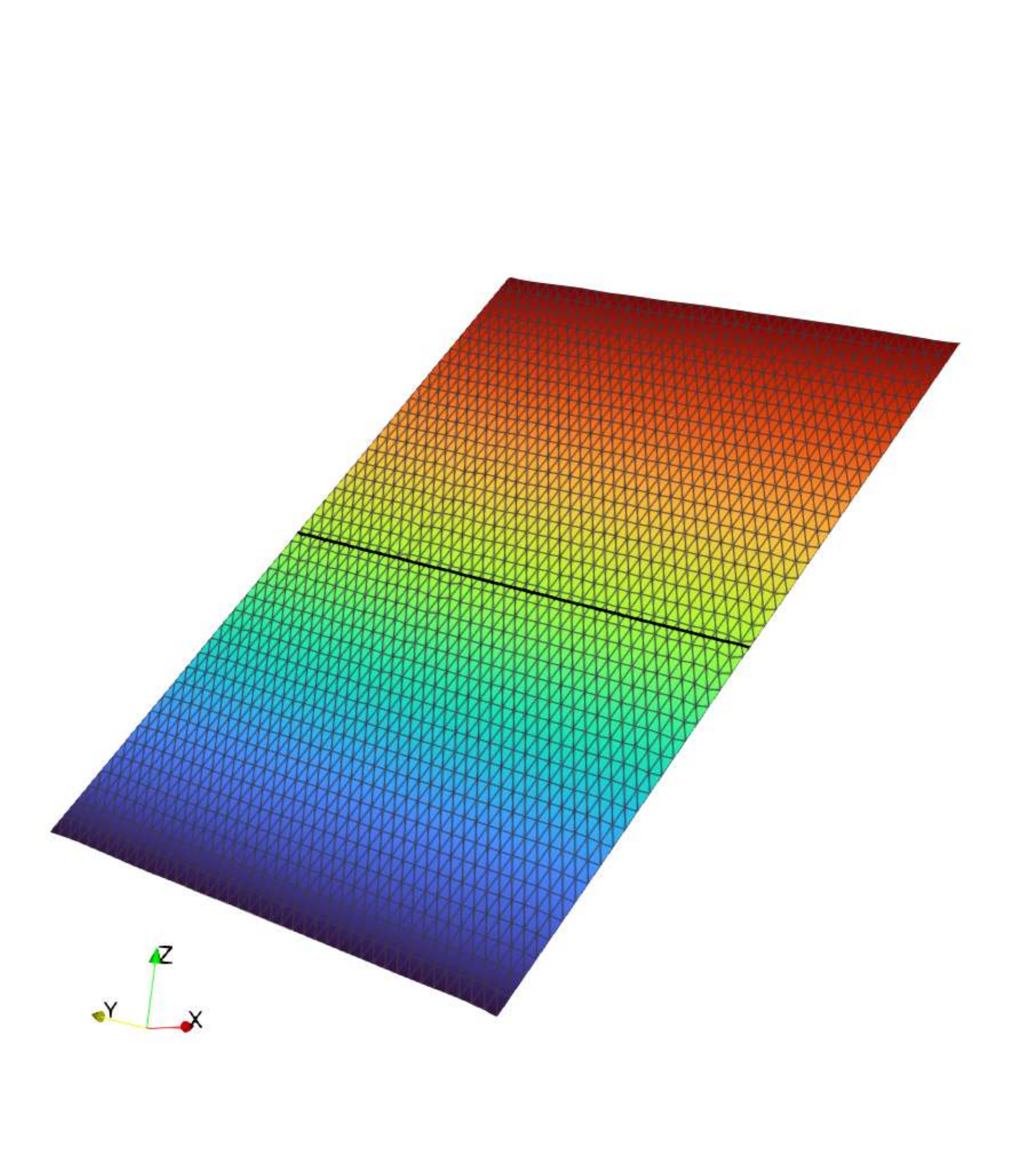}
        \caption{}\label{ls_reinit_PC_warp}
    \end{subfigure}
    \begin{subfigure}[b]{0.1\linewidth}\centering
         \includegraphics[width=0.9\textwidth]{legend_step_func.png}
    \end{subfigure}
    \caption{Warped level set for $\arctan$-problem with $\gamma_{D}=10$ and $h=\text{L}/41$: \subref{ls_init_PC_warp} level set function before reinitialization; \subref{ls_elliptic_warp} level set function after minimization based reinitialization; \subref{ls_reinit_PC_warp} level set function after P.C. reinitialization.}
    \label{fig:init_atan}
\end{figure}

\begin{figure}[!ht]
\centering
    \begin{subfigure}[b]{0.27\linewidth}\centering
        \includegraphics[width=0.9\textwidth]{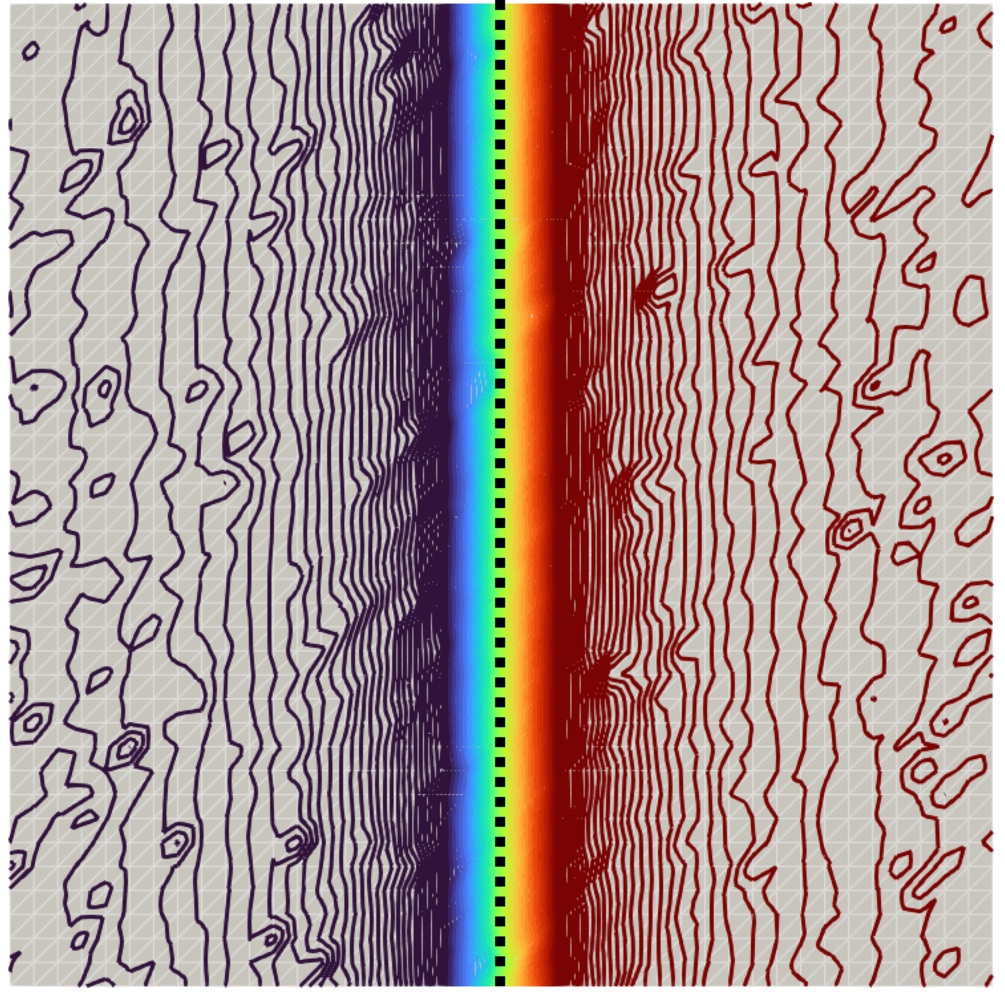}
        \caption{}\label{ls_init_PC_ls}
    \end{subfigure}
    \begin{subfigure}[b]{0.27\linewidth}\centering
         \includegraphics[width=0.9\textwidth]{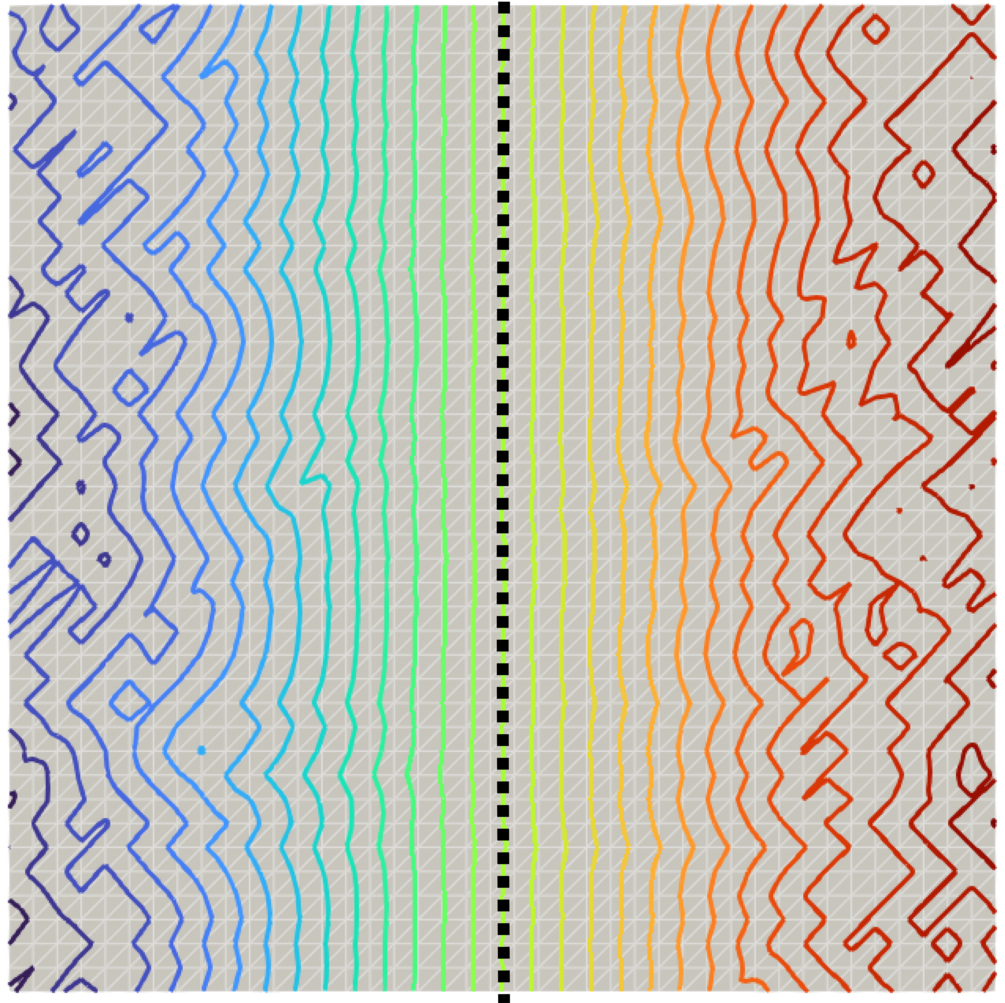}
        \caption{}\label{ls_elliptic_ls}
    \end{subfigure}
    \begin{subfigure}[b]{0.27\linewidth}\centering
         \includegraphics[width=0.9\textwidth]{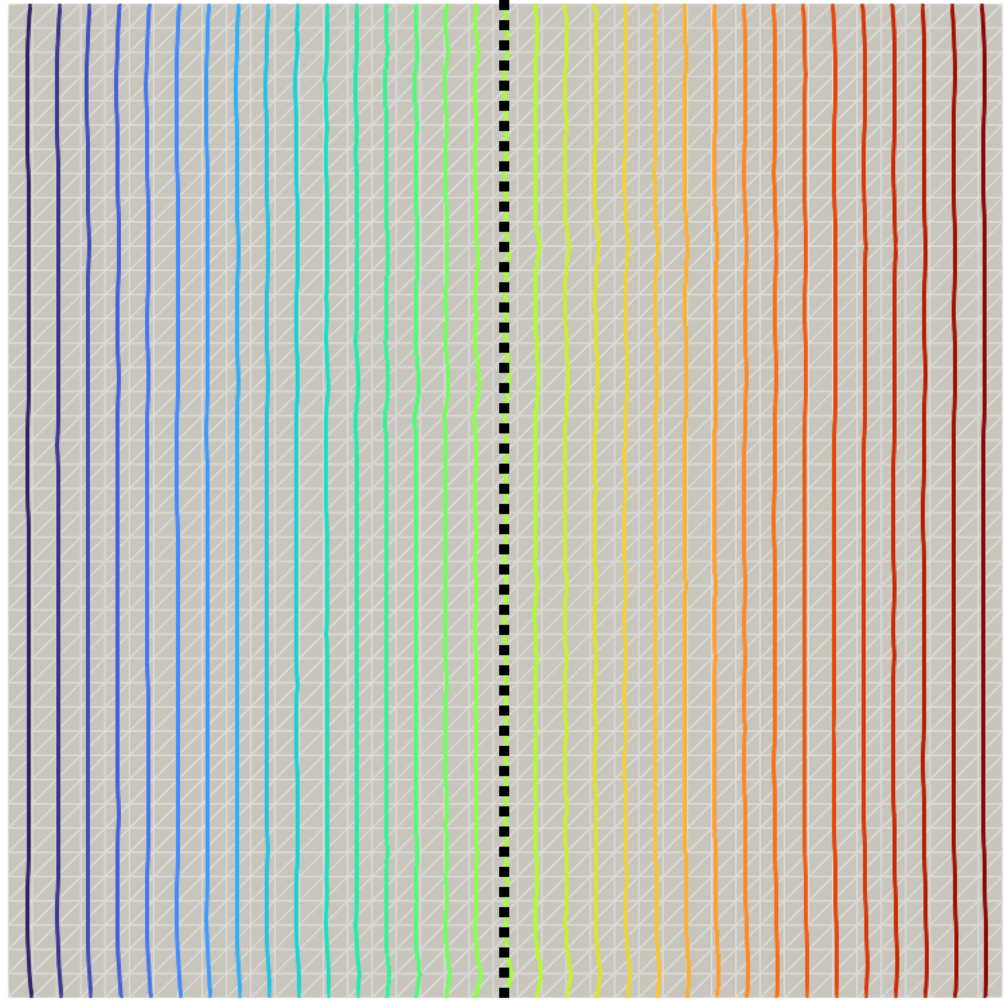}
        \caption{}\label{ls_reinit_PC_ls}
    \end{subfigure}
    \begin{subfigure}[b]{0.1\linewidth}\centering
         \includegraphics[width=0.9\textwidth]{legend_step_func.png}
    \end{subfigure}
    \caption{Perturbed $\arctan$ level set contours with $\gamma_{D}=10$ and $h=\text{L}/41$: \subref{stepfun_init_PC_warp} level set contours before reinitialization;  \subref{stepfun_corrector_PC_warp} level set contours after minimization based reinitialization;  \subref{stepfun_reinit_PC_warp} level set contours after P.C. reinitialization.}
    \label{level_set_atan_ls}
\end{figure}

\begin{figure}[H]
\centering
    \begin{subfigure}[b]{0.45\linewidth}\centering
        \includegraphics[width=1.\textwidth]{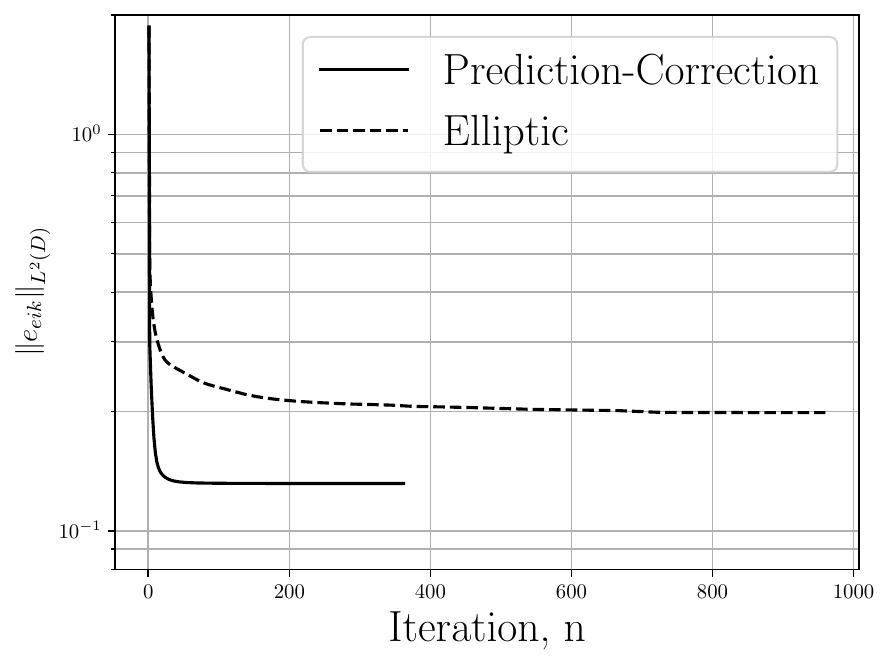}
        \caption{}
\label{error_atan_eik}
    \end{subfigure}
    \begin{subfigure}[b]{0.45\linewidth}\centering
         \includegraphics[width=1.\textwidth]{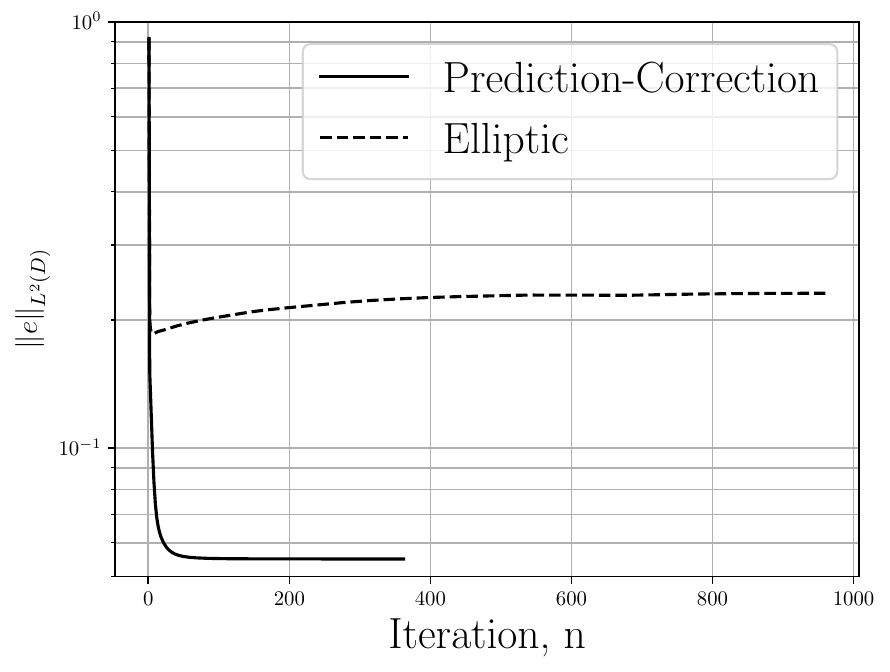}
        \caption{}\label{error_atan_e}
    \end{subfigure}
    \caption{Error evolution for  $\gamma_{D}=10$ and  $h=\text{L}/41$: \subref{error_atan_eik} error $\left\Vert e_{\text{eik}}\right\Vert_{L^{2}\left(D\right)}$; \subref{error_atan_e} error  $\left\Vert e\right\Vert_{L^{2}\left(D\right)}$.}
    \label{influence_gamma}
\end{figure}

To complete our study, we initialized a variant of the previous level set to assess the ability of the scheme to correct very irregular functions, containing areas with zero gradient norms and areas with very high gradient norms. 
The initialized level set is defined as:
\begin{equation}
    \phi_{\text{init}}(x,y) = \arctan\left( 10\left(x -0.5 \right)\right)+\xi(x)
\end{equation}
with $\xi$ a function of Gaussian small perturbation generated randomly.\\
This example highlights the relevance of the predictor problem for highly irregular initial level set functions. Indeed, as shown in Figures (\ref{error_atan_eik}) and (\ref{error_atan_e}), the errors obtained using the prediction-correction scheme are significantly lower than those resulting from solving the minimization problem alone. Looking at the level set function at convergence displayed in Figures~(\ref{fig:init_atan}) and (\ref{level_set_atan_ls}), we observe that the minimization problem alone is unable to smooth out significant irregularities in the vicinity of the domain boundaries while the predictor corrector problem yields a smooth approximation of the signed distance. Furthermore, it is observed that the elliptic scheme requires more iterations to converge to the solution. Finally, the error on $e$ decreases over iterations in the prediction-correction method, whereas this is not the case for the elliptic method as shown in Figure (\ref{error_atan_e}).

To illustrate the ability of our method to reinitialize level set with a complex interface, we'll look at an example next that features an irregular boundary. 

\subsection{Complex interface: star level set}\label{star_func}
\begin{figure}[!ht]
\centering
    \begin{subfigure}[b]{0.27\linewidth}\centering
        \includegraphics[width=0.9\textwidth]{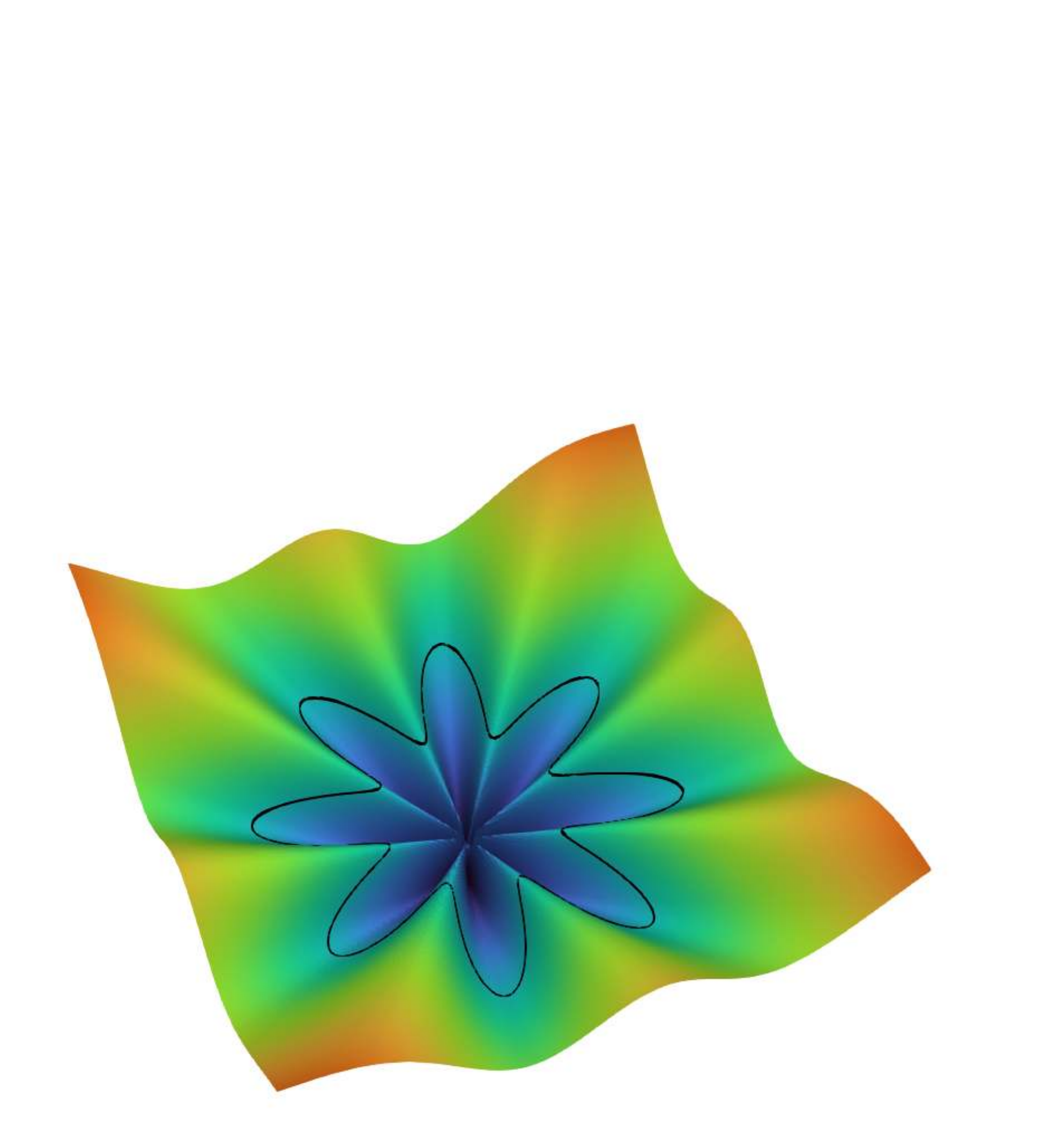}
        \caption{}\label{star_init_PC_warp}
    \end{subfigure}
    \begin{subfigure}[b]{0.27\linewidth}\centering
            \includegraphics[width=0.9\textwidth]{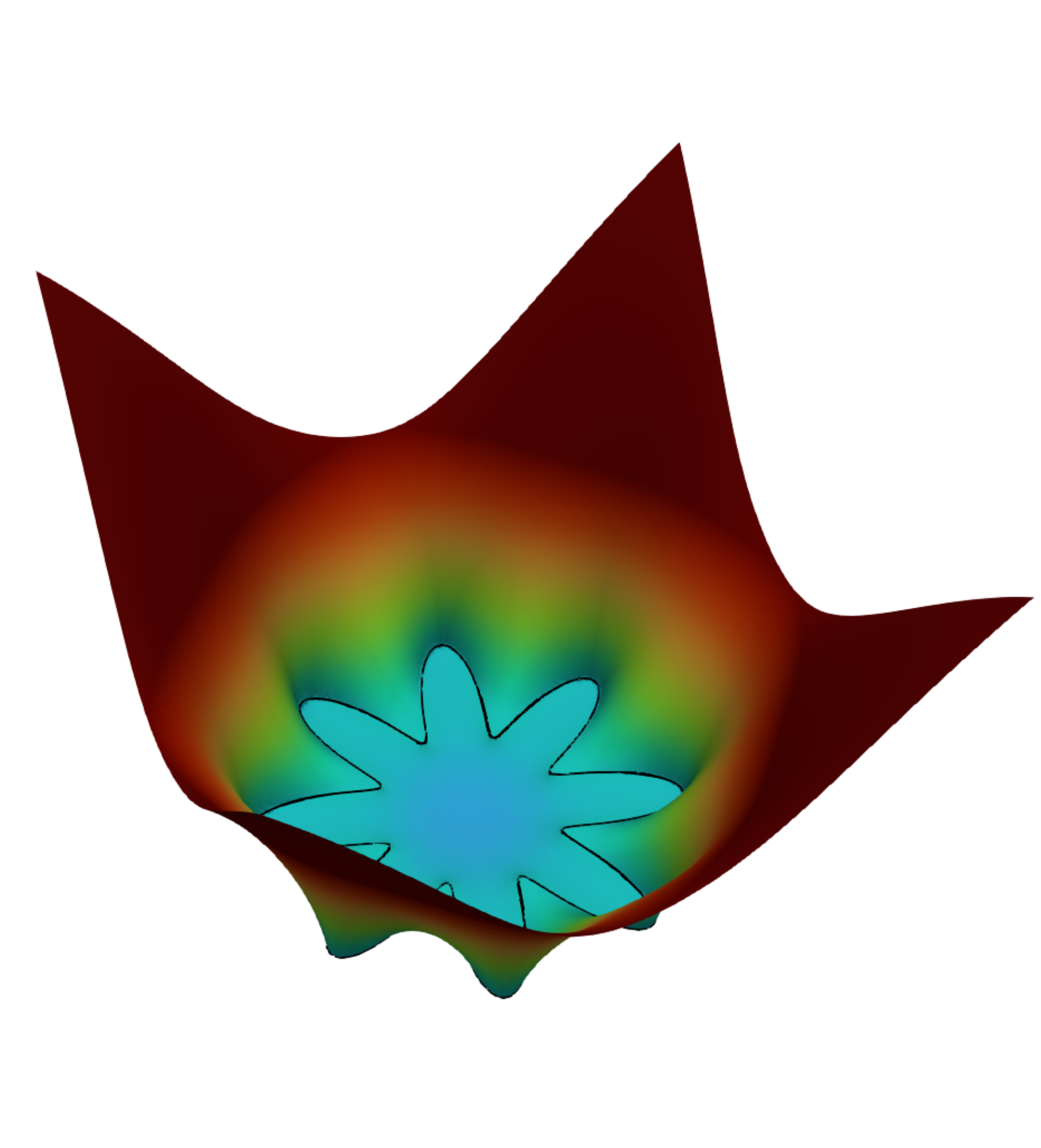}
        \caption{}\label{star_corrector_PC_warp}
    \end{subfigure}
    \begin{subfigure}[b]{0.27\linewidth}\centering
            \includegraphics[width=0.9\textwidth]{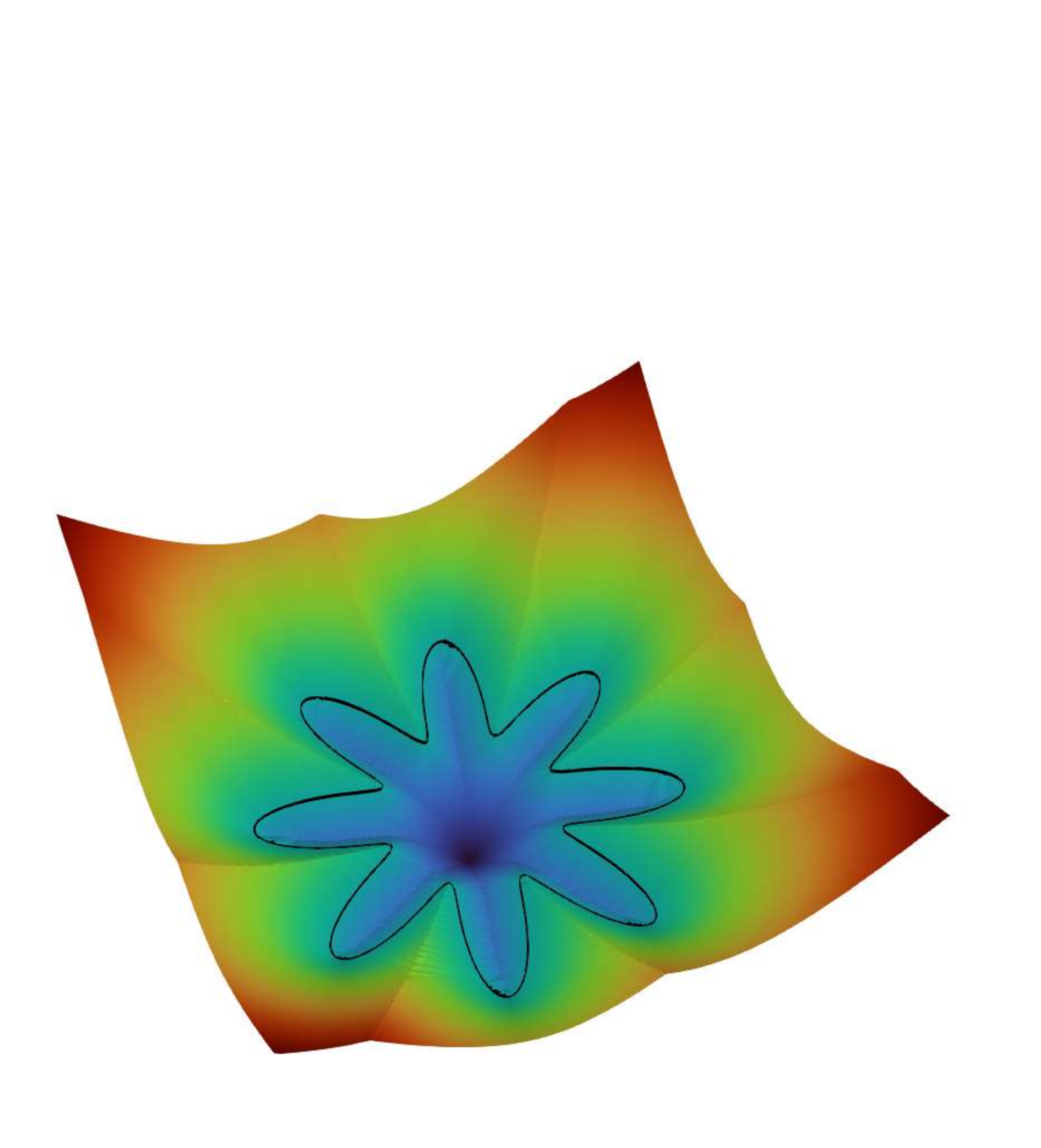}
        \caption{}\label{star_reinit_PC_warp}
    \end{subfigure}
    \begin{subfigure}[b]{0.1\linewidth}\centering
        \includegraphics[width=0.9\textwidth]{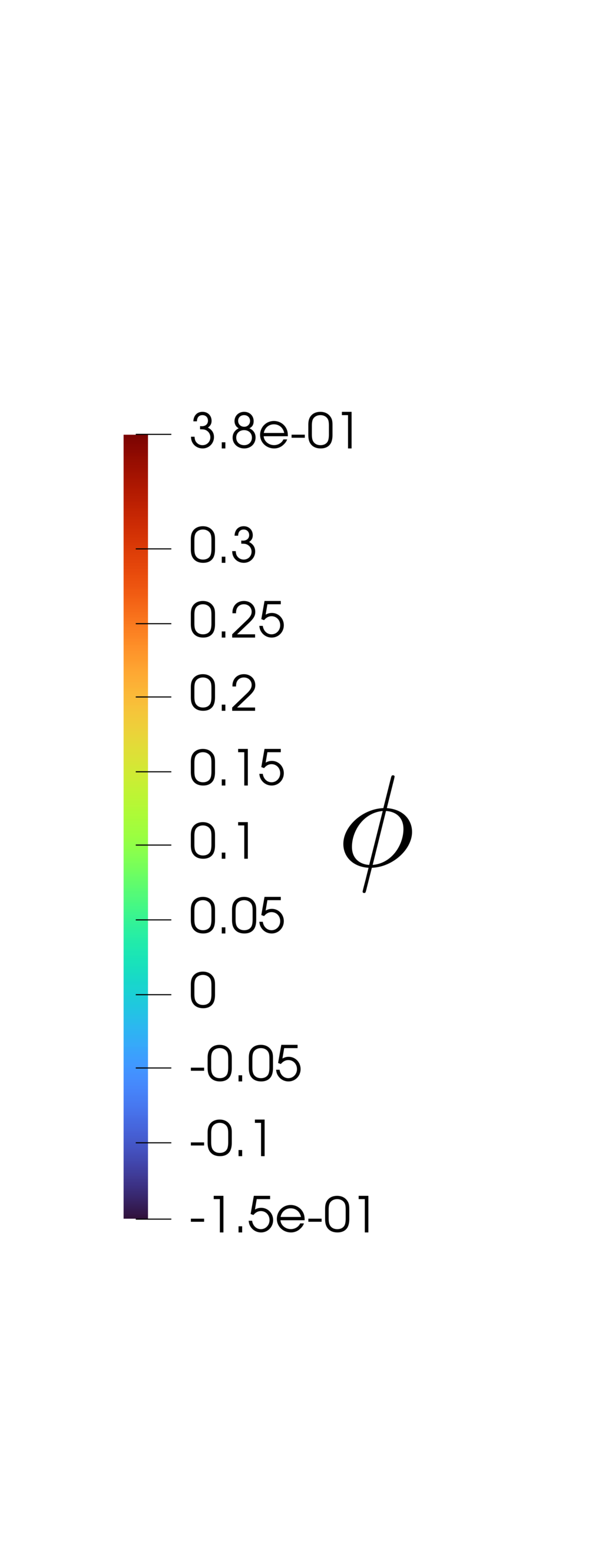}
    \end{subfigure}
        \caption{Warped star level set function with $\gamma_{D}=1000$ and $h=\text{L}/320$:\subref{star_init_PC_warp} before reinitialization; \subref{star_corrector_PC_warp} after predictor; \subref{star_reinit_PC_warp} after corrector reinitialization.}
    \label{warp_star}
\end{figure}

\begin{figure}[!ht]
    \centering
        \begin{subfigure}[b]{0.3\linewidth}\centering
            \includegraphics[width=0.9\textwidth]{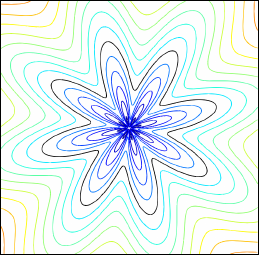}
            \caption{}\label{star_init_PC_ls}
        \end{subfigure}
        \begin{subfigure}[b]{0.3\linewidth}\centering
             \includegraphics[width=0.9\textwidth]{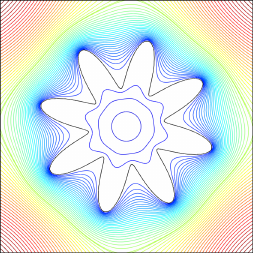}
            \caption{}\label{star_predictor_PC_ls}
        \end{subfigure}
        \begin{subfigure}[b]{0.3\linewidth}\centering
             \includegraphics[width=0.9\textwidth]{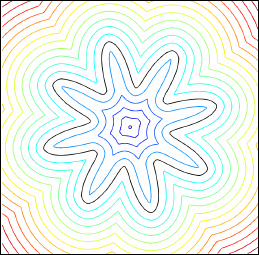}
            \caption{}\label{star_reinit_PC_ls}
        \end{subfigure}
        \caption{Star level set contours $\gamma_{D}=1000$ and $h=\text{L}/320$:\subref{star_init_PC_ls} before reinitialization; \subref{star_predictor_PC_ls} predictor problem; \subref{star_reinit_PC_ls} after corrector reinitialization.}
        \label{level_set_star}
    \end{figure}

In this section, we test our scheme on a star-shaped level set function (see Figures (\ref{star_init_PC_warp}) and (\ref{star_init_PC_ls}))  defined on the domain $D=\left[0,1\right]^{2}$ given by
\begin{equation}
    \phi_{\text{init}}\left(x,y\right)=\frac{1}{2}\left(\sqrt{\left(x-x_{0}\right)+\left(y-y_{0}\right)}-\frac{1}{10}\cos\left(n\arctan\left(\frac{y-y_{0}}{x-x_{0}}\right)+2\right)+\frac{1}{4}\right)
\end{equation}
with $\left(x_{0},y_{0}\right)$ the center of the star and $n=8$ the number of rays.
The distance function associated with this star-shaped form is particularly complex to compute. The interface exhibits large areas with high curvature and a lot of equidistant points along lines between the star rays for which the norm of the gradient is zero. The preservation of the geometry as well as the large number of equidistant lines makes this a challenging test problem. 

\subsection{ Simulation in three dimensions}
\begin{figure}[H]
    \centering
        \begin{subfigure}[b]{0.33\linewidth}\centering
            \includegraphics[width=0.9\textwidth]{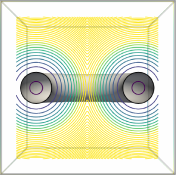}
            \caption{}\label{doughnut_init_ls}
        \end{subfigure}
        \begin{subfigure}[b]{0.33\linewidth}\centering
             \includegraphics[width=1\textwidth]{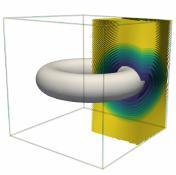}
            \caption{}\label{doughnut_init_warp}
        \end{subfigure}
        \begin{subfigure}[b]{0.1\linewidth}\centering
         \includegraphics[width=1\textwidth]{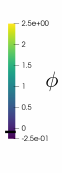}
    \end{subfigure}
         \caption{Doughnut level set contours before prediction-correction reinitialization: \subref{doughnut_init_ls} view of a 2D cross-section of the level lines with the interface in 3D dimensions;  \subref{doughnut_init_warp} view of a cross section of the level set. }
        \label{doughnut_init}
\end{figure}

\begin{figure}[H]
    \centering
        \begin{subfigure}[b]{0.33\linewidth}\centering
            \includegraphics[width=0.9\textwidth]{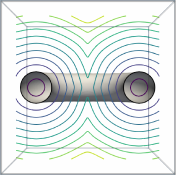}
            \caption{}\label{doughnut_reinit_ls}
        \end{subfigure}
        \begin{subfigure}[b]{0.33\linewidth}\centering
             \includegraphics[width=1\textwidth]{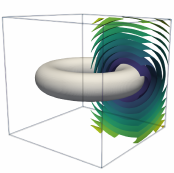}
            \caption{}\label{doughnut_reinit_warp}
        \end{subfigure}
        \begin{subfigure}[b]{0.1\linewidth}\centering
         \includegraphics[width=1\textwidth]{legend.pdf}
    \end{subfigure}
         \caption{Doughnut level set contours after prediction-correction reinitialization: \subref{doughnut_reinit_ls} view of a 2D cross-section of the level lines with the interface in 3D dimensions;  \subref{doughnut_reinit_warp} view of a cross section of the level set. }
        \label{doughnut_reinit}
\end{figure}

As our reinitialization scheme is purely based on PDEs and a cut integration of the interface local to each cell, our method can be parallelized easily and can hence be used for complex 3D geometries. In this section, we consider a doughnut-shaped level set function defined on the domain $D=\left[-\text{L}/2,\text{L}/2\right]^{3}$, with $\text{L}=2.5$ defined as follows:    
\begin{equation}\label{phi_init_doughnut}
    \phi_{\text{init}}\left(x,y,z\right) = \left(R-\sqrt{x^{2}-y^{2}}\right) + z^{2} - r^{2}
\end{equation}
with, $R = 1.8$ and $r=0.55$.
We use a tetrahedral mesh with size $h=\text{L}/101$, resulting in over two million unknowns.  Figure~\ref{doughnut_init} shows the initial level-set which is not a signed distance function as can be seen by the densely packed surfaces close to the zero contour surface.  After solving the predictor problem followed by 10 iterations of the corrector problem, Figure~\ref{doughnut_reinit} illustrates that the method gives a very good approximation of the signed distance function associated with the initial interface. The simulation was completed in 7 minutes and 54 seconds using 6 processors based on FEniCSx \cite{FEniCSx}, CutFEMx \cite{CutFEMx} as well as PETSc \cite{petsc-user-ref}.\\

\section{Conclusions}
In this article, we have presented a new finite element-based technique to compute the signed distance function starting from a level set function of general shape. Our method is based on a predictor-corrector formulation, where the predictor step is a linear diffusion problem with a discontinuous right hand side and the corrector step is a nonlinear diffusion problem based on the least squares minimization of the Eikonal equation. We presented both a finite element formulation for a mesh that fits to the zero contour line of the initial level set function as well as a cut finite element formulation that allows for a mesh that does not fit to the zero contour line.

We demonstrated that our predictor-corrector method is robust and converges to a signed distance function for a variety of initial level set functions, including smooth, discontinuous, and irregular functions. The method is also able to handle complex geometries in three dimensions and is easy to parallelize. We showed that the method is able to reinitialize the level set function with very few iterations, even for large meshes with millions of unknowns.

\section*{Acknowledgements}

This research was supported by the French Aerospace Lab ONERA under internal research funding. The authors gratefully acknowledge this support.


\end{document}